%% file: muse_HDFS.tex
\documentclass{aa}
\usepackage{paralist}
\usepackage{graphicx}
\usepackage{amssymb}
\usepackage{epstopdf}
\usepackage[varg]{txfonts}
\usepackage[LGRgreek]{mathastext}
\usepackage{lscape}
\usepackage{geometry}  
\usepackage{natbib}
\bibpunct{(}{)}{;}{a}{}{,} 

\def\lya{$Ly\alpha$}
\newcommand{\ciii}[1]{C\,\textsc{iii}]#1}
\newcommand{\oii}[1]{[O\,\textsc{ii}]#1}
\newcommand{\oiii}[1]{[O\,\textsc{iii}]#1}
\newcommand{\nii}[1]{[N\,\textsc{ii}]#1}

\newcommand{\mgii}[1]{Mg\,\textsc{ii}\,#1}
\newcommand{\neiii}[1]{[Ne\,\textsc{iii}]#1}
\newcommand{\ha}{\ensuremath{\mathrm{H}\alpha}}
\newcommand{\hb}{\ensuremath{\mathrm{H}\beta}}
\def\kms{km.s$^{-1}$}
\def\oiiib{[O\,{\sc iii}]$\lambda 5007$}
\def\hbeta{H$\beta$}
\def\imag{$I_{814}$}
\def\rmag{$V_{606}$}
\newcommand{\id}[1]{ID\##1}

\begin{document} 

 \title{The MUSE 3D view of the Hubble Deep Field South}

   \author{
           R. Bacon \inst{1}
        \and J. Brinchmann \inst{2}
        \and J. Richard  \inst{1}      
        \and T. Contini \inst{3,4} 
        \and  A. Drake  \inst{1} 
 	 \and M. Franx \inst{2}
	 \and S. Tacchella  \inst{5}
        \and J. Vernet  \inst{6}
   	 \and L. Wisotzki  \inst{7} 
	 \and J. Blaizot \inst{1}
	 \and N. Bouch\'e \inst{3,4} 
	 \and R. Bouwens \inst{2}
	 \and S. Cantalupo \inst{5}
	 \and C.M. Carollo \inst{5}
	 \and D. Carton \inst{2}
	 \and J. Caruana	\inst{7}
	 \and B. Cl\'ement \inst{1}	
	 \and S. Dreizler \inst{8}
	 \and B. Epinat \inst{3,4,9}	 
	 \and B. Guiderdoni \inst{1}
	 \and C. Herenz \inst{7} 
	 \and T.-O. Husser \inst{8}
	 \and S. Kamann \inst{8}
	 \and J. Kerutt \inst{7}
	 \and W. Kollatschny \inst{8}	
	 \and D. Krajnovic \inst{7}
	 \and S. Lilly \inst{5}
	 \and T. Martinsson \inst{2}
	 \and L. Michel-Dansac  \inst{1} 
	 \and V. Patricio \inst{1}
	 \and J. Schaye  \inst{2}
	 \and M. Shirazi \inst{5}	 
	 \and K. Soto \inst{5}
	 \and G. Soucail \inst{3,4}
	 \and M. Steinmetz \inst{7}
	 \and T. Urrutia \inst{7} 	 
	 \and P. Weilbacher \inst{7} 
	 \and T. de Zeeuw \inst{6,2}		   
    }

   \institute{
   CRAL, Observatoire de Lyon, CNRS, Universit\'e Lyon 1, 9 Avenue Ch. Andr\'e, F-69561 Saint Genis Laval Cedex, France
   \and Leiden Observatory, Leiden University, P.O. Box 9513, 2300 RA Leiden, The Netherlands 
   \and IRAP, Institut de Recherche en Astrophysique et Plan\'etologie, CNRS, 14, avenue Edouard Belin, F-31400 Toulouse, France 
   \and Universit\'e de Toulouse, UPS-OMP, Toulouse, France
   \and ETH Zurich, Institute of Astronomy, Wolfgang-Pauli-Str. 27, CH-8093 Zurich, Switzerland
   \and ESO, European Southern Observatory, Karl-Schwarzschild Str. 2, 85748 Garching bei Muenchen, Germany
   \and AIP, Leibniz-Institut f{\"u}r Astrophysik Potsdam, An der Sternwarte 16, D-14482 Potsdam, Germany
   \and AIG, Institut  f{\"u}r Astrophysik, Universit{\"a}t G{\"o}ttingen, Friedrich-Hund-Platz 1, D-37077 G{\"o}ttingen, Germany
   \and Aix Marseille Universit\'e, CNRS, LAM (Laboratoire d'Astrophysique de Marseille) UMR 7326, 13388, Marseille, France
   \\
   \\
    \email{roland.bacon@univ-lyon1.fr}
    }

   \date{Submitted November 27, 2014}
   
   \thanks{Based on observations made with ESO telescopes at the La Silla Paranal Observatory under program ID 60.A-9100(C).}

 
  \abstract 
 {We observed the Hubble Deep Field South with the new panoramic integral field spectrograph 
 MUSE that we built and just commissioned at the VLT. 
The data cube resulting from  27 hours of integration covers  one
$arcmin^2$ field of view at an unprecedented depth with a $1\sigma$ 
emission line surface brightness limit of
$1 \times 10^{-19} erg\,s^{-1}cm^{-2}\,arcsec^{-2}$, 
and contains $\sim$90,000 spectra. 
We present the combined  and calibrated data cube, and we perform a
first-pass analysis of the sources detected in the Hubble Deep Field
South imaging.
We measured the redshifts of 189 sources up to a magnitude \imag = 29.5,
increasing by more than an order of magnitude the number of
known spectroscopic redshifts in this field. We also discovered 26
\lya\ emitting galaxies which are not detected in the HST WFPC2 deep
broad band images. 

The intermediate spectral resolution of 2.3\AA\  allows us to separate resolved
asymmetric \lya\ emitters, \oii{3727} emitters, and \ciii{1908} emitters and 
the large instantaneous wavelength range of 4500 \AA\ helps to identify single
emission lines
such as \oiii{5007}, \hb, and \ha\ over a very large redshift range.
We also show how the three dimensional information of MUSE  helps to 
resolve sources which are confused at ground-based image quality.

Overall, secure identifications are provided for 83\% of the 227
emission line sources detected in the MUSE data cube and for 32\% of the 586 sources identified in the HST catalog of \cite{Casertano+2000}.
The overall redshift distribution is fairly flat to $z=6.3$, with a
reduction  between $z=1.5$ to 2.9, in the well-known redshift desert.
The field of view of MUSE also allowed us to detect 17 groups within the field.
We checked that the number counts of \oii{3727} and \lya\ emitters are roughly consistent with predictions from the literature.
Using two examples we demonstrate that MUSE is able to provide exquisite spatially resolved spectroscopic information on intermediate redshift galaxies present in the field.

This unique data set can be used for a large range of follow-up studies. 
We release the data cube, the associated products, and the source
catalogue with redshifts, spectra and emission line fluxes.
}


   \maketitle
%


\section{Introduction}
The Hubble deep fields North and South \citep[see e.g.][]{Williams+1996, Ferguson+2000, Beckwith+2006} are still among the deepest images ever obtained in the optical/infrared, providing broad band photometry for sources up to V$\sim$30 . Coupled with extensive multi-wavelength follow-up campaigns they, and the subsequent Hubble Ultra Deep Field, have been instrumental in improving our understanding of galaxy formation and evolution in the distant Universe. 

Deep, broad-band, photometric surveys provide a wealth of information on the galaxy population, such as  galaxy morphology, stellar masses and photometric redshifts. Taken together this can be used to study the formation and evolution of the Hubble sequence \citep[e.g.][]{Mortlock+2013, Lee+2013}, the change in galaxy sizes with time \citep[e.g.][]{vanderWel+2014, Carollo+2013}, and the evolution of the stellar mass function with redshift \citep[e.g.][]{Muzzin+2013, Ilbert+2013}.  But photometric information alone gives only a limited view of the Universe: essential physical information such as the kinematic state of the galaxies and their heavy element content require spectroscopic observations. Furthermore, while photometric redshifts work well on average for many bright galaxies \citep[e.g.][]{Ilbert+2009}, they have insufficient precision for environmental studies, are occasionally completely wrong, and their performance on very faint galaxies is not well known.

Ideally, one would like to obtain spectroscopy for all sources at the same depth than the broad band photometry. However  current technology does not mach these requirements.
For example, the VIMOS ultra-deep survey \citep{Lefevre+2014} which is today the largest spectroscopic deep survey with 10,000 observed galaxies in 1 deg$^2$, is in general limited to R$\sim$25 and only 10\% of the galaxies detected in the Hubble Deep Field North and South are that bright. 

Another fundamental limitation, when using multi-object spectrographs,
is the need to pre-select a sample based on broad band imaging. Even
if it were feasible to target all objects found in the Hubble Deep
Fields WFPC2 deep images (i.e.\ $\sim 6000$ objects), 
the sample would still not include all galaxies with high equivalent width emission lines, even though determining redshifts for these would be relatively easy.
For example, faint low mass galaxies with high star
formation rate at high redshifts 
may not have an optical counter-part even in very deep HST broad-band imaging,
although their emission lines arising in their star-forming interstellar medium might be detectable spectroscopically.

Long slit observations are not a good alternative because of the limited field of view and other technical limitations due to slits such as
 unknown slit light losses, loss of positional information perpendicular to the slit, possible velocity errors, etc.
For example, \cite{Rauch+2008}   performed a long slit integration
of 92 hours with FORS2 at the VLT. They targeted  the redshift range of 2.67-3.75 and 
their observations went very deep, with emission line surface brightness limit ($1\sigma$) depth of $8.1\times 10^{-20} \mathrm{erg}\,
\mathrm{sec}^{-1} \,\mathrm{cm}^{-2}\,\mathrm{arcsec}^{-2}$. However, this performance was obtained with a field of view of 0.25 $arcmin^2$ and only one spatial dimension, limiting the usefulness of this technique for follow-up surveys of Hubble deep fields.

To overcome some of these intrinsic limitations, a large, sensitive
integral field spectrograph is required. It must be sensitive and stable enough to be 
able to reach a depth commensurate  to that of the Hubble deep fields, while at 
the same time having a high spatial resolution, large multiplex, spectral coverage,
and good spectral resolution.  This was at the origin of the Multi Unit
Spectroscopic Explorer (MUSE) project to build a panoramic integral
field spectrograph for the VLT (\citealt{Bacon+2010}; Bacon et
al. in prep).  The commissioning of MUSE 
on the VLT was completed in August 2014 after
development by a consortium of 7 European institutes: CRAL (lead),
AIG, AIP, ETH-Zurich, IRAP, NOVA/Leiden Observatory and ESO. The instrument
has a field of view of $1 \times 1$ arcmin$^2$ sampled at 0.2 arcsec, an
excellent image quality (limited by the 0.2 arcsec sampling), a large simultaneous spectral range
(4650-9300\,\AA), a medium spectral resolution (R $\simeq$ 3000) and a
very high throughput (35\% end-to-end including telescope at 7000~\AA).

Although MUSE is a general purpose instrument and has a wide range of applications (see \citealt{Bacon+2014} for a few illustrations), it has been, from the very start of the project in 2001, designed and optimised for performing deep field observations. Some preliminary measurements performed during the first commissioning runs had convinced us that MUSE was able to reach its combination of high throughput and excellent image quality. However it is only by performing a very long integration on a deep field that one can assess the ultimate performance of the instrument. This was therefore one key goal of the final commissioning run of ten nights in dark time late July 2014. The Hubble Deep Field South (HDFS) which was observable during the second half of the nights for a few hours, although at relatively high airmass, was selected as the ideal target to validate the performance of MUSE, the observing strategy required for deep fields to limit systematic uncertainties and to test the data reduction software.

The HDFS was observed  with the Hubble Space Telescope in 1998
\citep{Williams+2000}. The WFPC2 observations \citep{Casertano+2000}
reach a $10 \sigma$ limiting AB magnitude in the F606W filter (hereafter \rmag) at  28.3 and
27.7 in the F814W filter (hereafter \imag). The  field  was one of the first
to obtain very deep Near-IR multi-wavelength observations (e.g.,
Labb\'e et al 2003).
But, contrarily to the Hubble Ultra Deep Field \citep{Beckwith+2006}
which had very extensive spectroscopic follow-up observations, the
main follow-up efforts on the HDFS have been imaging surveys
(e.g. \citealt{Labbe+2003, Labbe+2005}).

In the present paper we show that the use of a wide-field, highly
sensitive IFU provides a very powerful technique to target such deep
HST fields, allowing a measurement of $\sim$200 redshifts per arcmin$^2$, $\sim$30 of which are not detected in the HDF continuum images. 
We present the first deep observations taken with the MUSE
spectrograph, and we determine a first list of secure redshifts
based on emission-line and absorption-line features. We show that the
high spectral resolution, large wavelength range, and three dimensional
nature of the data help to disentangle confused galaxies and to
identify emission lines securely.

The paper is organised as follows: the observations, data
reduction and a first assessment of instrument performance are described in the next two sections. In sections
\ref{sect:ident} and \ref{sect:census}, we proceed with the source
identifications and perform a first census of the field content.  The
resulting redshift distribution and global properties of the detected
objects are presented in section \ref{sect:redshifts}. Examples of kinematics analysis
performed on two spatially  resolved galaxies is given in section
\ref{sect:resolved}. 
Comparison with the current generation of deep spectroscopic surveys and conclusions are given in the last section. 
The data release is described in the Appendices.

\section{Observations}
\label{sect:obs}
\begin{figure} \resizebox{\hsize}{!}{\includegraphics{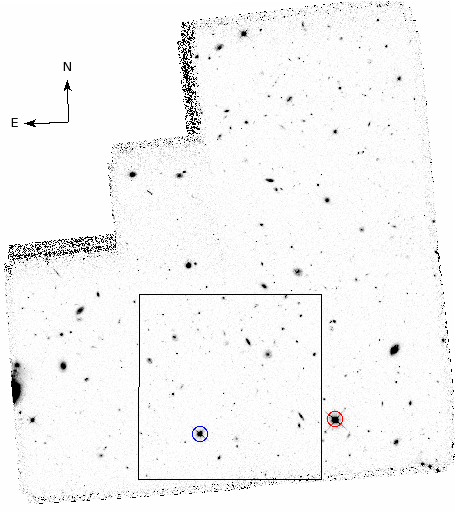}}
\caption{Location of the MUSE field of view within the HDFS F814W image. The star used in the slow guiding system is indicated in red and  the brightest star in the field (R=19.6) in blue.}
\label{fig:field}
\end{figure}

The HDFS was observed during six nights in July 25--29, 31 and August 2, 3 2014 of the last commissioning run of MUSE. The $1\times1$ arcmin$^2$ MUSE  field was centered at 
$\alpha = 22h32'55.64''$, $\delta =-60^o33'47''$. This location was selected in order to have one bright star in the Slow Guiding System (SGS) area and another bright star in the field of view (Fig \ref{fig:field}). We used the nominal wavelength range (4750-9300\,\AA) and performed a series of exposures of 30 minutes each.
The spectrograph was rotated by  90$^o$ after each integration, and
the observations were dithered using random offsets within a 3 arcsec box.
This scheme ensures that
most objects will move from one channel 
\footnote{The field of view of MUSE is first split in 24 ``channels'', each channel is then split again in 48 ``slices" by the corresponding image slicer.}
 to a completely different one while at the same time minimizing the field loss. This is, however, not true
for the objects that fall near the rotation centre.

In addition to the standard set of calibrations, we obtained a
flat field each hour during the night. These 
single flat field exposures, referred to as attached flats in the following, are used to correct for the small
illumination variations caused by temperature variations during the
night. 
The Slow
Guiding System was activated for all exposures using a bright R=19.2
star located in the SGS field. The SGS also gives an accurate real-time estimate of
the seeing which was good for most of the nights (0.5--0.9 arcsec). Note that the
values given by the SGS are much closer to the seeing achieved in the science exposure
than the values given by the DIMM seeing monitor.  An
astrometric solution was derived using an off-center field of a
globular cluster with HST data. A set of spectrophotometric standard
stars was also observed when the conditions were photometric.

In total, 60 exposures of 30 mn integration time were obtained. A
few exposures were obtained in cloudy conditions and were
discarded. One exposure was lost due to an unexpected VLT guide star
change in the middle of the exposure. The remaining number of
exposures was 54, with  a total integration time of 27 hours. One of these exposures 
was offset by approximately half the field of view to test the performance of the SGS guiding on a faint galaxy.

\section{Data Reduction and Performance analysis}
\label{sect:datared}
\subsection{Data reduction process}
The data were reduced with version 0.90 of the MUSE standard
pipeline. The pipeline will be described in detail in  Weilbacher
et al (in prep.)\footnote{A short description is also given in
  \cite{Weilbacher+2012}}. We summarize the main steps to produce the fully reduced data cube:
\begin{enumerate}
\item Bias, arcs and flat field master calibration solutions were created using a set of standard calibration exposures obtained each night. 
\item Bias images were subtracted from each science
  frame. Given its low value, the dark current ($\sim1\, e^-.hour^{-1}$, that is $0.5\, e^-$ per exposure) was neglected.
 Next, the science frames were flat-fielded using the master flat field and renormalized using the  attached flat field as an illumination correction. An additional flat-field correction was performed using the twilight sky exposures to correct for the difference between sky and calibration unit illumination. The result of this process is a large table (hereafter called a pixel-table) for each science frame. This table contains all pixel values corrected for bias and flat-field and their location on the detector. A geometrical calibration and the wavelength calibration  solution were used to transform the detector coordinate positions to wavelengths in {\AA}ngstr{\"o}m and focal plane spatial coordinates.
\item The astrometric solution was then applied.  The flux calibration was
  obtained from observations of the spectrophotometric standard star
  Feige 110 obtained on August 3, 2014. We verified that the system
  response curve was stable between the photometric nights with a measured scatter below 0.2\% rms.
The response
  curve was smoothed with spline functions to remove  high
  frequency fluctuations left by the reduction. Bright sky lines were
  used to make small corrections to  
 the wavelength solution obtained from the master arc.
  All these operations have been done at the
  pixel-table level to avoid unnecessary interpolation. The formal noise 
  was also calculated at each step. 
\item To correct for the small shifts introduced by the derotator
  wobble between exposures, we fitted a Gaussian function to the
  brightest star in the reconstructed white-light image of the field. The astrometric solution of the
  pixel-tables of all  exposures was normalized to the HST catalog
  coordinate of the star ($\alpha = 22h37'57.0''$, $\delta =-60^o34'06''$) 
  The fit to the star  also provides an accurate measurement
  of the seeing of each exposure. The average Gaussian white-light FWHM value for the 54
  exposures is $0.77 \pm 0.15$ arcsec. We also derived the total
  flux of the reference star by simple aperture photometry, the maximum variation
  among all retained exposures is 2.4\%.
\item To reduce systematic mean zero-flux level offsets between slices,
  we implemented a
  non-standard self-calibration process. From a first reconstructed white-light
  image produced by the merging of all exposures, we derived a mask to
  mask out all bright continuum objects present in the field of
  view. For each exposure, we first computed the median flux over all
  wavelengths and the non-masked spatial coordinates. Next we
  calculated the median value for all  slices, and we 
  applied an additive correction to each slice to bring all slices to
  the same median value.
   This process very effectively  removed residual offsets between
   slices.
\item A data cube was produced from each pixel-table using a 3D drizzle interpolation process which include sigma-clipping to reject outliers such as cosmic rays. All data cubes were sampled to a common grid in view of the final combination (0\farcs2 $\times$ 0\farcs2 $\times$ 1.25 \AA).
\item We used the software ZAP (Soto et al, in prep.) to subtract the
  sky signal from each of the individual exposures.  ZAP operates by first subtracting a baseline sky level, found by calculating the median per spectral plane, leaving any residuals due to variations in the line spread function and system response. The code then uses principal component analysis on the data cube to calculate the eigenspectra and eigenvalues that characterize these residuals, and determines the minimal number of eigenspectra that can reconstruct the residual emission features in the data cube.
\item The 54 data cubes were then merged in a single data cube using
  $5\sigma$ sigma-clipped mean.  The variance for each combined volume pixel or 'voxel' was computed as the variance derived from the comparison of the N individual exposures divided by $N-1$, where N is the number of voxels left after the sigma-clipping.  This variance data cube is saved as an additional extension of the combined data cube.
In addition an exposure map data cube which counts the number of exposures used for the combination of each voxel was also saved. 
\item Telluric absorption from H$_2$O and O$_2$ molecules was fitted to the spectrum of a white dwarf found in the field ($\alpha = 22h32'58.77''$, $\delta =-60^o33'23.52''$) using the \texttt{molecfit} software described in Smette et al, submitted. and Kausch et al, submitted. In the fitting process, the line spread function was adjusted using a wavelength dependent Gaussian kernel. The resulting transmission correction was then globally applied to the final datacube and variance estimation.
\end{enumerate}

The result of this process is a fully calibrated data cube of 3 Gb size with spectra in the first extension and the variance estimate in the second extension, as well as an exposure cube of 1.5 Gb) size giving the number of exposures used for each voxel. 

\subsection{Reconstructed white-light image and Point Spread Functions}

\begin{figure} \resizebox{\hsize}{!}{\includegraphics{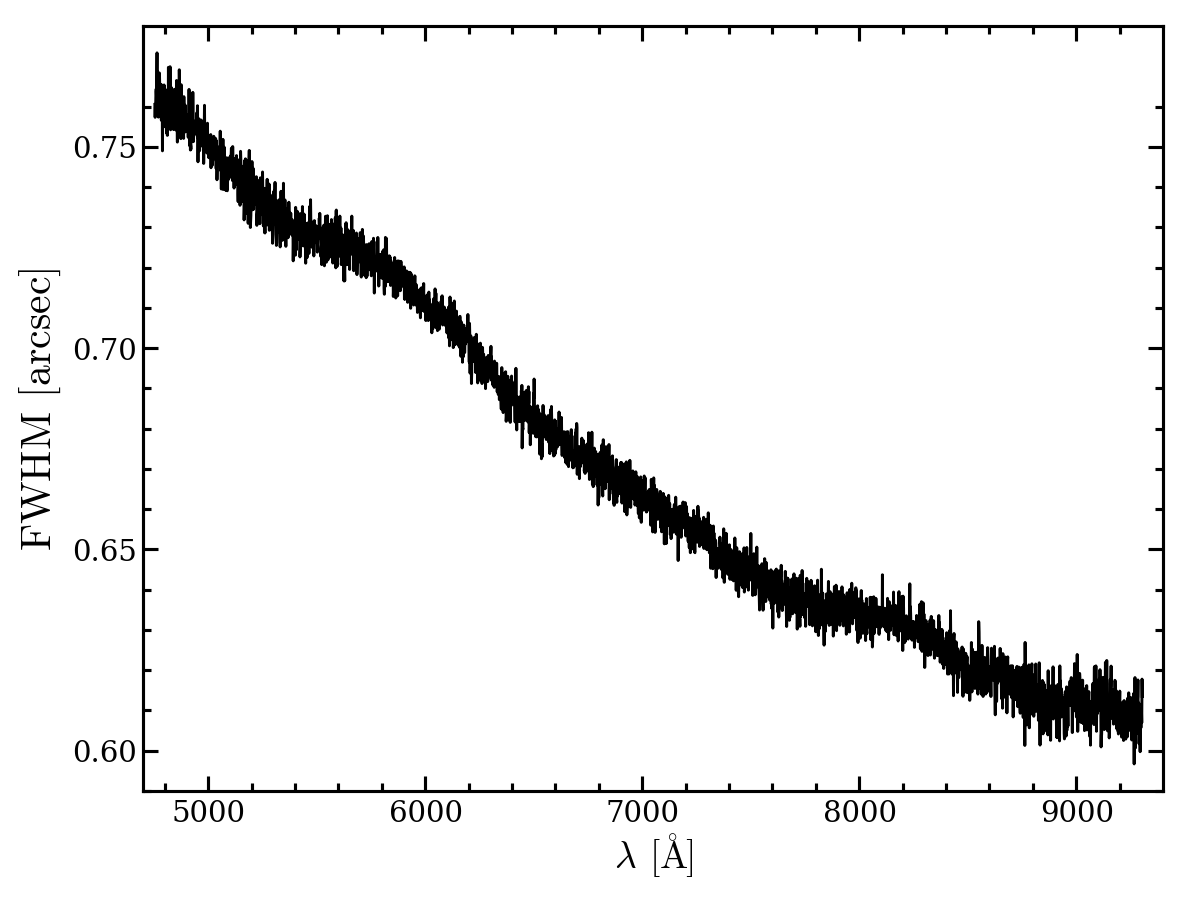}}
\caption{The derived FWHM as function of wavelength from the Moffat fit  to the brightest star in the field.}
\label{fig:fwhm_lbda}
\end{figure}

The image quality was assessed using a Moffat fit of the reference star as a function of wavelength. The PSF shape is circular  with a fitted Moffat $\beta$ parameter of 2.6 and a FWHM of 0.66 arcsec at 7000\AA. While  $\beta$  is almost constant with wavelength, the FWHM shows the expected trend with wavelength decreasing from 0.76 arcsec in the blue to 0.61 arcsec in the red (Fig.~\ref{fig:fwhm_lbda}). Note that the FWHM derived from the MOFFAT model is systematically 20\% lower than the Gaussian approximation.

The  spectral Line Spread Function (LSF) was measured on arc calibration frames. We obtain an average value of $2.1 \pm 0.2$ pixels which translates into a spectral resolution of R $3000 \pm 100$ at 7000\AA. A precise measurement of the LSF shape is difficult because it is partially under sampled. In the present case this uncertainty is not problematic because the spectral features of the identified objects are generally broader than the LSF.

\begin{figure}
\centering
 \resizebox{\hsize}{!}{\includegraphics{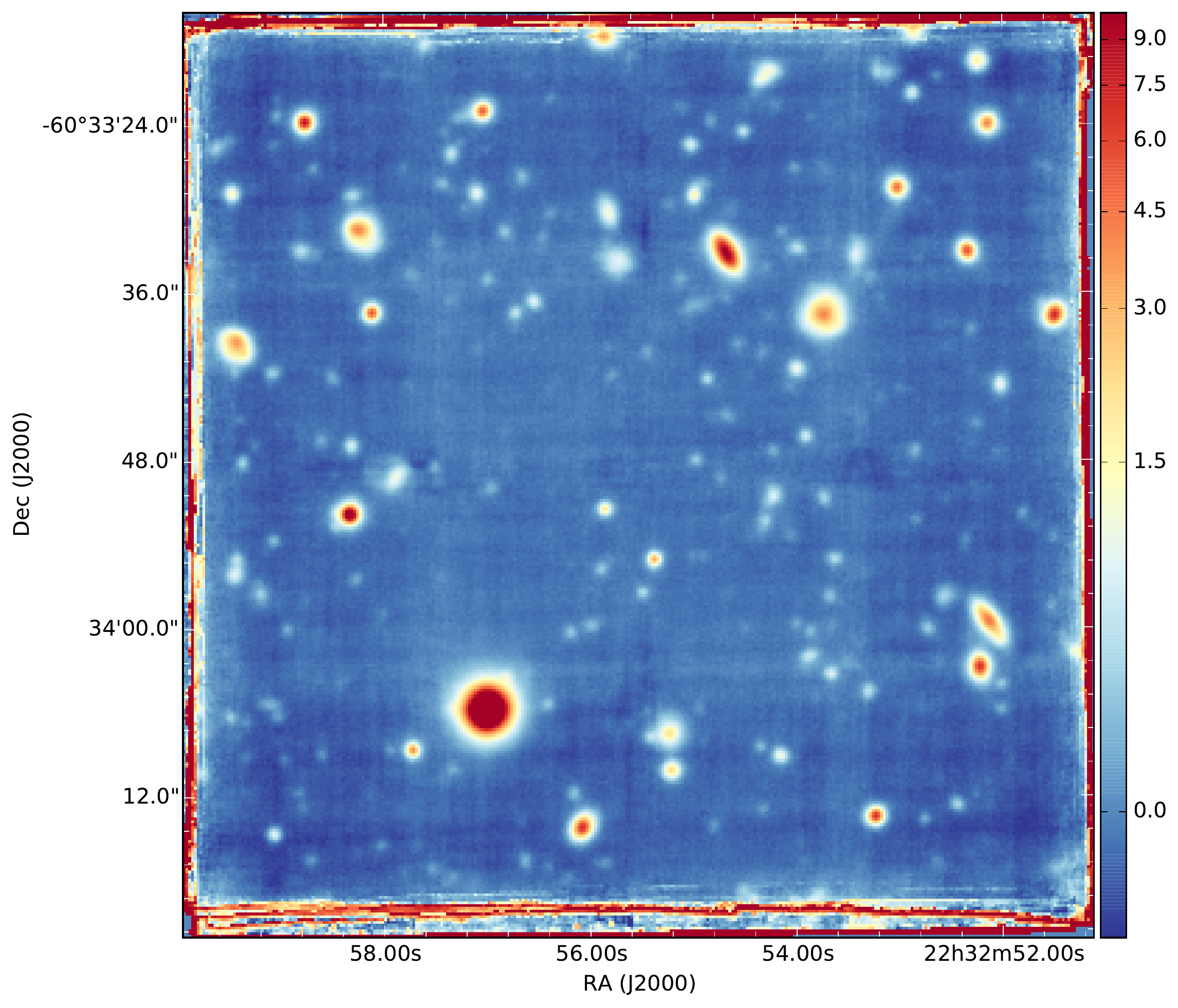}}
\caption{Reconstructed white-light image of the combined exposures. The flux scale shown on the right is in
$10^{-21} erg\,s^{-1}cm^{-2}\AA^{-1}pixel^{-1} $. Orientation is North(up)-East(left)}
\label{fig:white-light}
\end{figure}

A simple average over all wavelengths gave the reconstructed white
light image (Fig. \ref{fig:white-light}). Inspection of this
image reveals numerous objects, mostly galaxies.  
The astrometric accuracy, derived by comparison  with the \cite{Casertano+2000} catalogue, is $\sim 0.1$ arcsec. At lower flux levels, $F \sim 2 \times 10^{-21} erg\,s^{-1}cm^{-2}\AA^{-1}pixel^{-1} $, some residuals of the instrument channel splitting can be seen
in the reconstructed white-light image in the form of a series of vertical and horizontal stripes. 

\subsection{Signal to noise ratios}
\label{subsec:lw_snr}

Characterising the noise in a MUSE datacube is not trivial, as each voxel of the cube is interpolated not only spatially, but also in the spectral domain, and each may inherit flux from just one up to $\sim$30 original CCD pixels. While readout and photon noise are formally propagated by the MUSE pipeline as pixel variances, correlations between two neighbouring spatial pixels cause these predicted variances to be systematically too low.  Furthermore, the degree of correlation between two neighbouring spatial pixels varies substantially across the field (and with wavelength), on spatial scales comparable or larger than real astronomical objects in a deep field such as the HDFS. A correct error propagation also accounting for the covariances between pixels is theoretically possible, but given the size of a MUSE dataset this is unfortunately prohibitive with current computing resources. We therefore have to find other ways to estimate the ``true'' noise in the data.

For the purpose of this paper we focus on faint and relatively small sources, and we therefore neglect the contribution of individual objects to the photon noise. In addition to readout and sky photon noise, unresolved low-level systematics can produce noise-like modulations of the data, especially when varying rapidly with position and wavelength. Such effects are certainly still present in MUSE data, e.g.\ due to the residual channel and slice splitting pattern already mentioned in Sect.~\ref{sect:datared}. Another issue are sky subtraction residuals of bright night sky emission lines which are highly nonuniform across the MUSE field of view. Future versions of the data reduction pipeline will improve on these features, but for the moment we simply absorb them into an ``effective noise'' budget.

In the HST image of the HDFS we selected visually a set of 100 ``blank sky'' locations free from any continuum or known emission line sources. These locations were distributed widely over the MUSE field of view, avoiding the outskirts of the brightest stars and galaxies. We extracted spectra through circular apertures from the sky-subtracted cube which consequently should have an expectation value of zero at all wavelengths. Estimates of the effective noise were obtained in two different ways: (A) By measuring the standard deviation inside of spectral windows selected such that no significant sky lines are contained; 
(B) by measuring the standard deviation of aperture-integrated fluxes between the 100 locations, as a function of wavelengths. Method (A) directly reproduces the ``noisiness'' of extracted faint-source spectra, but cannot provide an estimate of the noise for all wavelengths. Method (B) captures the residual systematics also of sky emission line subtraction, but may somewhat overestimate the ``noisiness'' of actual spectra. We nevertheless used the latter method as a conservative approach to construct new ``effective'' pixel variances that are spatially constant and vary only in wavelength.
Overall, the effective noise is higher by a factor of $\sim$1.4 than the local pixel-to-pixel standard deviations, and by a factor of $\sim$1.6 higher than the average propagated readout and photon noise. Close to the wavelengths of night sky emission lines, these factors may get considerably higher, mainly because of the increased residuals.

The median effective noise per spatial and spectral pixel for the HDFS datacube is then $9\times 10^{-21}$~erg\,s$^{-1}$cm$^{-2}$\AA$^{-1}$, outside of sky lines. For an emission line extending over 5~\AA\ (i.e. 4 spectral pixels), we derive a $1 \sigma$ emission line surface brightness limit of 
$1\times 10^{-19} \mathrm{erg}\,\mathrm{sec}^{-1} \,\mathrm{cm}^{-2}\,\mathrm{arcsec}^{-2}$.

An interesting comparison can be made with the \cite{Rauch+2008} deep long slit integration. 
In 92 hours, they reached a $1\sigma$ depth of $8.1\times 10^{-20} \mathrm{erg}\,
\mathrm{sec}^{-1} \,\mathrm{cm}^{-2}\,\mathrm{arcsec}^{-2}$, again summed over 5~\AA.
\citeauthor{Rauch+2008} covered the wavelength range 4457--5776~\AA, which is 3.4$\times$ times smaller than the MUSE wavelength range, and they cover an area (0.25 arcmin$^2$) which is four times smaller. 
Folding in the ratio of exposure times and the differences in achieved flux limits, the MUSE HDFS datacube is then in total over 32 times more effective for a blind search of emission line galaxies than the FORS2 observation.

From the limiting flux surface brightness one can also derive the limiting flux for a point source. This is, however, more complex because it depends of the seeing and the extraction method. A simple approximation is to use fix aperture. For a 1 arcsec diameter aperture, we measured a light loss of 40\% at 7000~\AA\ for the brightest star in the MUSE field. Using this value, 
we derive an emission line limiting flux at $5 \sigma$
of $3 \times 10^{-19} \mathrm{erg}\,\mathrm{sec}^{-1} \,\mathrm{cm}^{-2}$ for a point source within a 1 arcsec aperture.

\begin{figure}
\includegraphics[width=\linewidth]{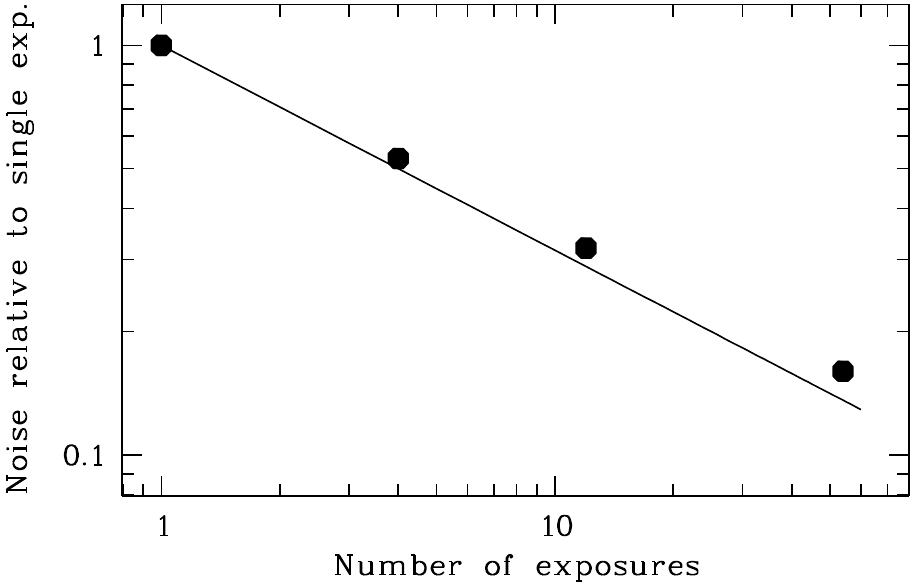} 
\caption{Overall scaling of pixel noise measured in the cubes as a function of the number of exposures combined, relative to the noise in a single exposure. The solid line represents the ideal $1\,/\sqrt{n}$ behaviour.} 
\label{fig:lw_noisescaling}
\end{figure}

In order to have a better understanding of the contribution of systematics to the noise budget, we also investigated the scaling of the noise with exposure time. Taking the effective noise in a single 30~min exposure as unit reference, we also measured the noise in coadded cubes of 4, 12, and the full set of 54 exposures. While systematic residuals are also expected to decrease because of the rotational and spatial dithering, they would probably not scale with $1\,/\sqrt{n}$  as perfect random noise. The result of this exercise is shown in Fig.~\ref{fig:lw_noisescaling}, which demonstrates that a significant deviation from $1\,/\sqrt{n}$ is detected, but that it is quite moderate (factor $\sim 1.2$ for the full HDFS cube with $n = 54$). Even without reducing the systematics, adding more exposures would make the existing HDFS dataset even deeper.

\section{Source identification and redshift determination}
\label{sect:ident}

The three-dimensional nature of the MUSE observations presents unique challenges,
while at the same time offering multiple ways to extract spectra and to determine and confirm redshifts.  We have found that constructing 1D and 2D projections of the sources is essential to ascertain redshifts for the fainter sources. In particular we have constructed 1D spectra and for each tentative emission line we construct a continuum subtracted narrow-band image over this line, typically with a width of 10\AA, and only if this produces a coherent image of the source do we consider the emission line real.  Likewise, 2D spectra can be useful additional tools for understanding the spatial distribution of emission.

The extraction of spectra from a very deep data cube can be challenging since the ground based seeing acts to blend sources. Here too the construction of 2D images can help disentangling sources that otherwise would be blended. For this step the existence of deep HST images is very helpful to interpret the results.

This method does not lend itself to absorption line redshifts. In this case we examine the spatial variation of possible absorption line features by extracting spectra at various spatial positions. A real absorption line should be seen in multiple spectra across the galaxy.

\subsection{Redshift determination of continuum detected objects}
\label{sect:continuum}
We extracted subcubes around each object in the \citet{Casertano+2000} catalogue that fell
within the FoV sampled by the observations.  We defined our spectrum extraction aperture by running SExtractor \citep{Bertin+1996}, version 2.19.5  on the reconstructed white-light images. In the case when no object was clearly detected in the white-light image, a simple circular extraction aperture with diameter 1.4" was used. When a redshift was determined, we constructed narrow-band images around  Ly-$\alpha$, \ciii{1909}, \oii{3727}, \hb, \oiii{5007}, and \ha, whenever that line fell within the wavelength range of MUSE, and ran SExtractor on these as well.
The union of the emission-line and the white-light segmentation maps define our object mask. The SExtractor segmentation map was also used to provide a sky mask. The object and sky masks were inspected and manually adjusted when necessary to mask out nearby sources and to avoid edges. 

The local sky residual spectrum is constructed by averaging the spectra within the sky mask. The object spectrum was constructed by summing the spectra within the object mask, subtracting off the average sky spectrum  in each spatial pixel (spaxel). We postpone the optimal extraction of spectra \citep[e.g.][]{Horne1986} to future work as this is not essential for the present paper. Note also that we do not account for the wavelength variation of the PSF in our extraction --- this is a significant concern for optimal extraction but for the straight summation we found by testing on stars within the datacube that the effect is minor for extraction apertures as large as ours. 

\begin{figure*}
   \includegraphics[width=184mm]{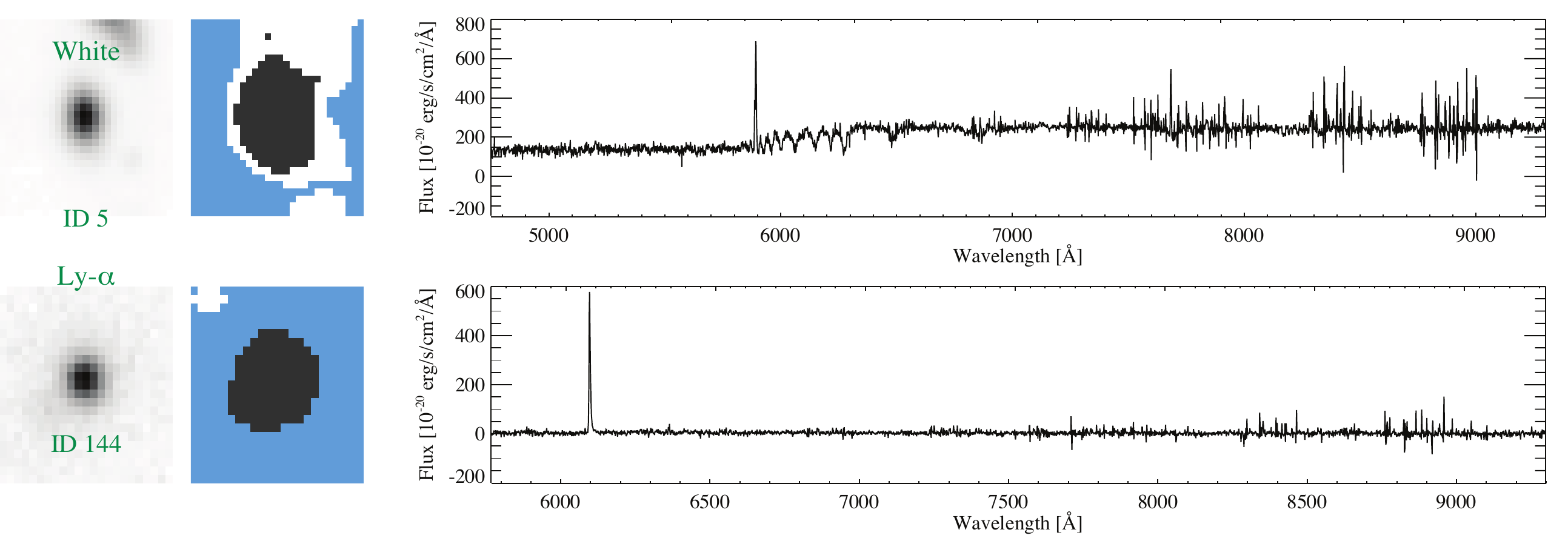} 
   \caption{An example of the extraction process for objects 5 and 144, at $z=0.58$ and $z=4.02$ respectively. In the top row the left panel shows the MUSE reconstructed white-light image, while the left panel in the bottom row shows the \lya\ narrow-band image. The middle panels show the object and sky masks, with the object aperture shown in black and the sky aperture in blue. Part of the final extracted spectrum is shown in the right-hand panel on each row.}
   \label{fig:extraction-example}
\end{figure*}

An example of the process can be seen in Figure~\ref{fig:extraction-example}. The top row shows the process for a $z\approx 0.58$ galaxy with the white-light image shown in the left-most column, and the bottom row the same for a \lya-emitter at $z=4.02$ with the \lya-narrow-band image shown in the left-most column. The region used for the local sky subtraction is shown in blue in the middle column, while the object mask is shown in black.

The resulting spectra were inspected manually and emission lines and absorption features were identified by comparison to template spectra when necessary.  In general an emission line redshift was considered acceptable if a feature consistent with an emission line was seen in the 1D spectrum and coherent spatial feature was seen in several wavelength planes over this emission line. In some cases mild smoothing of the spectrum and/or the cube was used to verify the reality of the emission line. In the case of absorption line spectra several absorption features were required to determine a redshift. 

In many cases this process gives highly secure redshifts, with multiple lines detected in 72 galaxies and 8 stars. We assign them a Confidence $=3$. In general the identification of single line redshifts is considerably less challenging than in surveys carried out with low spectral resolution. The \oii{3726,3729} and \ciii{1907,1909} doublets are in most cases easily resolved and the characteristic asymmetric shape of \lya\ is easily identified.  In these cases we assign a Confidence $=2$ for single-line redshifts with high signal-to-noise. In a number of cases we do see unresolved \oii{3726,3729} --- in these cases we still have a secure redshift from Balmer absorption lines and/or \neiii{3869} --- but these lines are unresolved due to velocity broadening and we do not expect to see this behaviour in spectra of very faint galaxies. The other likely cases of single line redshifts with a \textit{symmetric} line profile are: \ha-emitters with undetected \nii{6584} and no accompanying strong \oiii{5007}; \oiii{5007} emitters with undetectable \oiii{4959} and \hb; and \lya-emitters with symmetric line profiles.

To distinguish between these alternatives, we make use of two
methods. The first is to examine the continuum shape of the spectra.
For the brighter objects breaks in the spectra can be used to separate
between the various redshift solutions.  The second method is to
check the spectrum at the location of any other possible line, and
to extract  narrow-band images for all possible strong lines
--- this is very useful for \oiii{5007} and \ha\ emitters, and can exclude or
confirm low redshift solutions. If this process does not lead to a secure redshift, we assign a confidence $0$ to these sources.  
The redshift confidence assignments is summarized below:
\begin{compactitem}
  \setlength\itemsep{0em}
  \item 0: No secure or unique redshift determination possible 
  \item 1: Redshift likely to be correct but generally based on only one feature
  \item 2: Redshift secure, but based on one feature
  \item 3: Redshift secure, based on several features
\end{compactitem}

\smallskip
In the case of overlapping sources we do not attempt to optimally extract the spectrum of each source, leaving this to future papers. In at least four cases we see two sets of emission lines in the extracted spectrum and we are unable to extract each spectrum separately. Despite this we are still generally able to associate a redshift to a particular object in the HST catalog by looking at the distribution of light in narrow-band images. We also use these narrow band images to identify cases where a strong emission line in a nearby object contaminates the spectrum of a galaxy.

\subsection{Identification  of line emitters without continuum}
In parallel to the extraction of continuum-selected objects, we also searched for  
sources detected only by their line emission. Two approaches were used: a visual inspection of the 
MUSE data cube, and a systematic search using automatic detection tools.

Two of the authors (JR, TC) visually explored the data cube over its full wavelength 
range in search of sources appearing only in a narrow wavelength range, typically 4-5 
wavelength planes ($\sim$6-7 Angstroms), and seemingly extended over a number of pixels 
at least the size of the seeing disk. We then carefully inspected the extracted line 
profile around this region to assess the reality of the line.

Any visual inspection has obvious limits and we also employed more automatic tools for 
identifying sources dominated by emission lines. One such tool is based on SExtractor 
\citep{Bertin+1996} which was run on narrow-band images produced by averaging 
each wavelength plane of the cube with the 4 closest wavelength planes, adopting a 
weighting scheme which follows the profile on an emission line with a velocity $\sigma=100$ km/s. 
This procedure was performed accross the full wavelength range of the cube to enhance 
the detection of single emission lines (see also \citealt{Richard+2014}). All 
SExtractor catalogs obtained from each narrow-band image were merged and compared with 
the continuum estimation from the white light image to select emission lines. 

A different approach was used in the LSDCat software (Herenz,
in prep.), which 
was specifically designed to search for line emitters not associated with continuum 
sources in the MUSE data cube. 
The algorithm is based on matched filtering (e.g \citealt{Das1991}):
by cross-correlating the data cube with a template that resembles the
expected 3D-signal of an emission line, the signal-to-noise ratio of a
faint emission line is maximized. The optimal template
for the search of compact emission line objects is a function that
resembles the seeing PSF in the spatial domain and a general emission
line shape in the spectral domain. In practice we use a 3D template
that is a combination of a 2D Moffat profile with a 1D Gaussian spectral
line. The 2D Moffat parameters is taken from the bright star fit 
(see Fig. \ref{fig:fwhm_lbda} in Sect. \ref{sect:datared})
and the FWHM of the Gaussian is fixed to 300 km/s in
velocity space. To remove continuum signal, we median-filter the
data cube in the spectral direction and subtract this cube 
from the original cube.  In the
following cross-correlation operation the variances are propagated
accordingly, and the final result is a data cube that contains a formal
detection significance for the template in each cube element (voxel). Thresholding is
performed on this data cube, where regions with neighboring voxels
above the detection threshold are counted as one object and a catalog of
positions $(x,y,\lambda)$ of those detections is created.  
To limit the number of false detections due to unaccounted systematics from
sky-subtraction residuals in the redder part of the data cube, a detection threshold of 10$\sigma$ 
was used. The candidate sources were then visually inspected by 3 authors (CH, JK \& JB).
This process results in the addition of 6 new identifications that had escaped the previous inspections.


All the catalogs of MUSE line emitters described above were
cross-correlated with the \cite{Casertano+2000} catalog of continuum sources presented in
Section \ref{sect:continuum}. A few of the emission lines
were associated with continuum sources based on their projected
distance in the plane of the sky.
Isolated emission lines not
associated with HST continuum sources were treated as separate entries in the
final catalog and their spectra were extracted blindly at the locations
of the line emission. The spectral extraction procedure was identical
to the one described in Section \ref{sect:continuum}.

The very large majority of line emitters not associated with HST continuum detections show a clear 
isolated line with an asymmetric profile, which we associate with \lya\ emission. 
In most case this is corroborated by the absence of other strong lines (except possibly 
\ciii{} emission at $2.9<z<4$) and the absence of a resolved doublet (which would be expected 
in case of \oii{} emission). 

\subsection{Line flux measurements}
\label{sec:flux_measurements}

We measure emission line fluxes in the spectra using the \texttt{platefit} code described by \citet{Tremonti+2004} and \citet{Brinchmann+2004} and used for the MPA-JHU catalogue of galaxy parameters from the SDSS\footnote{\texttt{http://www.mpa-garching.mpg.de/SDSS}}. This fits the stellar spectrum using a non-negative least-squares combination of theoretical spectra broadened to match the convolution of the velocity dispersion and the instrumental resolution. It then fits Gaussian profiles to emission lines in the residual spectrum.  Because of its asymmetric shape, this is clearly not optimal for \lya\ and we describe a more rigorous line flux measurement for this line in section~\ref{subsec:lw_nc} below.

The signal-to-noise in most spectra is insufficient for a good determination of the stellar velocity dispersion, so for the majority of galaxies we have assumed a fixed intrinsic velocity dispersion of 80 \kms. This resulted in good fits to the continuum spectrum for most galaxies and changing this to 250 \kms changes forbidden line fluxes by less than 2\%, while for Balmer lines the effect is $<5$\% for those galaxies for which we cannot measure a velocity dispersion. These changes are always smaller than the formal flux uncertainty and we do not consider these further here.

The emission lines are fit jointly with a single width in velocity space and a single velocity offset relative to the continuum redshift. Both the \oii{3726,3729} and \ciii{1906,1908} doublets are fit separately so the line ratios can be used to determine electron density. We postpone this calculation for future work.

\subsection{Comparison between MUSE and published spectroscopic redshifts}
\label{subsec:zspec}
Several studies have provided spectroscopic redshifts for sources in the HDFS and its flanking fields:
\begin{itemize}
\item \cite{Sawicki+2003} presented spectroscopic redshifts for 97 $z<1$ galaxies with FORS2 at the VLT. Their initial galaxy sample was selected based on photometric redshifts, $z_{phot}\la0.9$ and their resulting catalog is biased towards $z\sim0.5$ galaxies. The spectral resolution ($R\sim2500-3500$) was sufficient to resolve the \oii{3727} doublet, enabling a secure spectroscopic redshift determination. The typical accuracy that they quote for their spectroscopic redshifts is $\delta z = 0.0003$.
\item \cite{Rigopoulou+2005} followed up 100 galaxies with FORS1 on the VLT, and measured accurate redshifts for 50 objects. Redshifts were determined based on emission lines (usually \oii{3727}) or, in a few cases, absorption features such as the CaII H, K lines. The redshift range of the spectroscopically-detected sample is $0.6-1.2$, with a median redshift of 1.13. These redshifts agree well with the \cite{Sawicki+2003} estimates for sources in common between the two samples. 
\item \cite{Iwata+2005} presented VLT/FORS2 spectroscopic observations of galaxies at $z\sim3$; these were selected to have $2.5 < z_{phot} < 4$ based on HST/WFPC2 photometry combined with  deep near-infrared images obtained with ISAAC at the VLT by \cite{Labbe+2003}. They firmly identified five new  redshifts as well as two additional tentative redshifts of $z\sim3$ galaxies.
\item \cite{Glazebrook+2006} produced  53 additional extragalactic redshifts in the range $0<z<1.4$ with the AAT Low Dispersion Survey Spectrograph  by targeting 200 objects with $R > 23$. 
\item Finally \cite{Wuyts+2009} used a variety of optical spectrographs on $8-10$ m class telescopes (LRIS and DEIMOS at Keck Telescope, FORS2 at the  VLT and GMOS at Gemini South) to measure redshifts for 64 optically faint distant red galaxies. 
\end{itemize}

In total these long term efforts have provided a few hundred spectroscopic redshifts. They are, however, distributed over a much larger area than the proper HDFS deep imaging WFPC2 field, which has only $\sim 88$ confirmed spectroscopic redshifts. In the MUSE field itself, which covers 20\% of the WFPC2 area, we found 18 sources in common (see Table \ref{table:specz} in Appendix~\ref{sect:catalog}). As shown in Fig.~\ref{fig:redshifts}, most of the 18 sources cover the bright part (\imag $< 24$) of the MUSE redshift-magnitude distribution. 

Generally speaking there is an excellent agreement between the different  redshift estimates over the entire redshift range covered by the 18 sources.  Only one major discrepancy is detected: \id{2} (HDFS J223258.30-603351.7), which \cite{Glazebrook+2006} estimated  to be at redshift 0.7063 while MUSE reveals it to be a star.  These authors gave however a low confidence grading of $\sim50\%$ to their identification. After excluding this object, the agreement is indeed excellent with a normalized median difference of $\left< \Delta z/(1+z) \right> = 0.00007$ between MUSE and the literature estimates. 

\begin{figure}
\resizebox{\hsize}{!}{\includegraphics{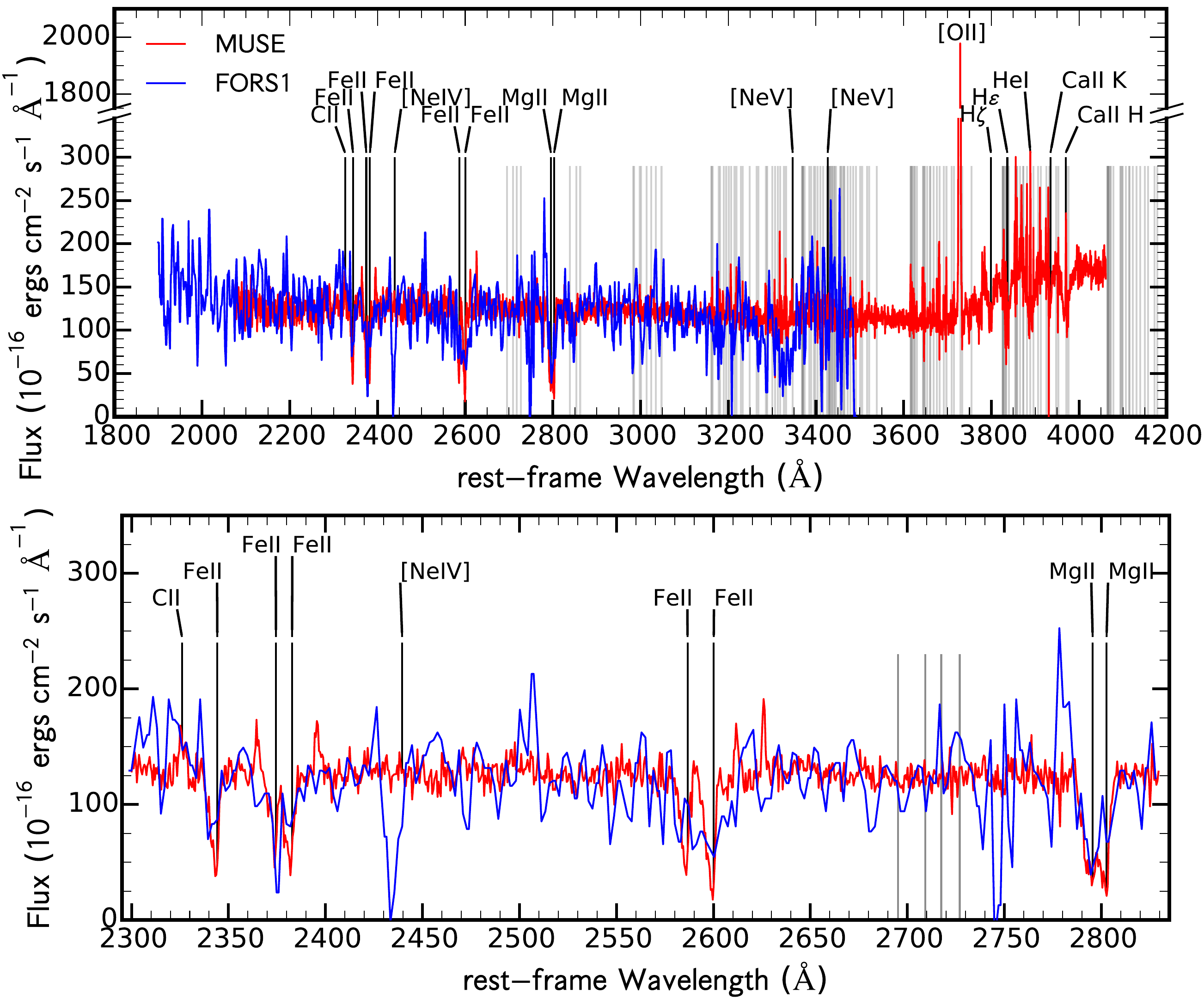}}
\caption{Comparison between  the MUSE (red) and FORS1 spectra of \citealt{Rigopoulou+2005} (blue) for the galaxy \id{13}, a strong \oii\ emitter at $z_{MUSE}=1.2902$. The strongest spectral features are indicated in black. The grey lines show the position of the sky lines. The upper panel shows the entire spectra;   the lower panel shows a zoom on the rest-frame 2300-2800$\AA$ region, which contains strong MgII and FeII absorption features.} 
\label{fig:spec_ID_13}
\end{figure}

\begin{figure}
\resizebox{\hsize}{!}{\includegraphics{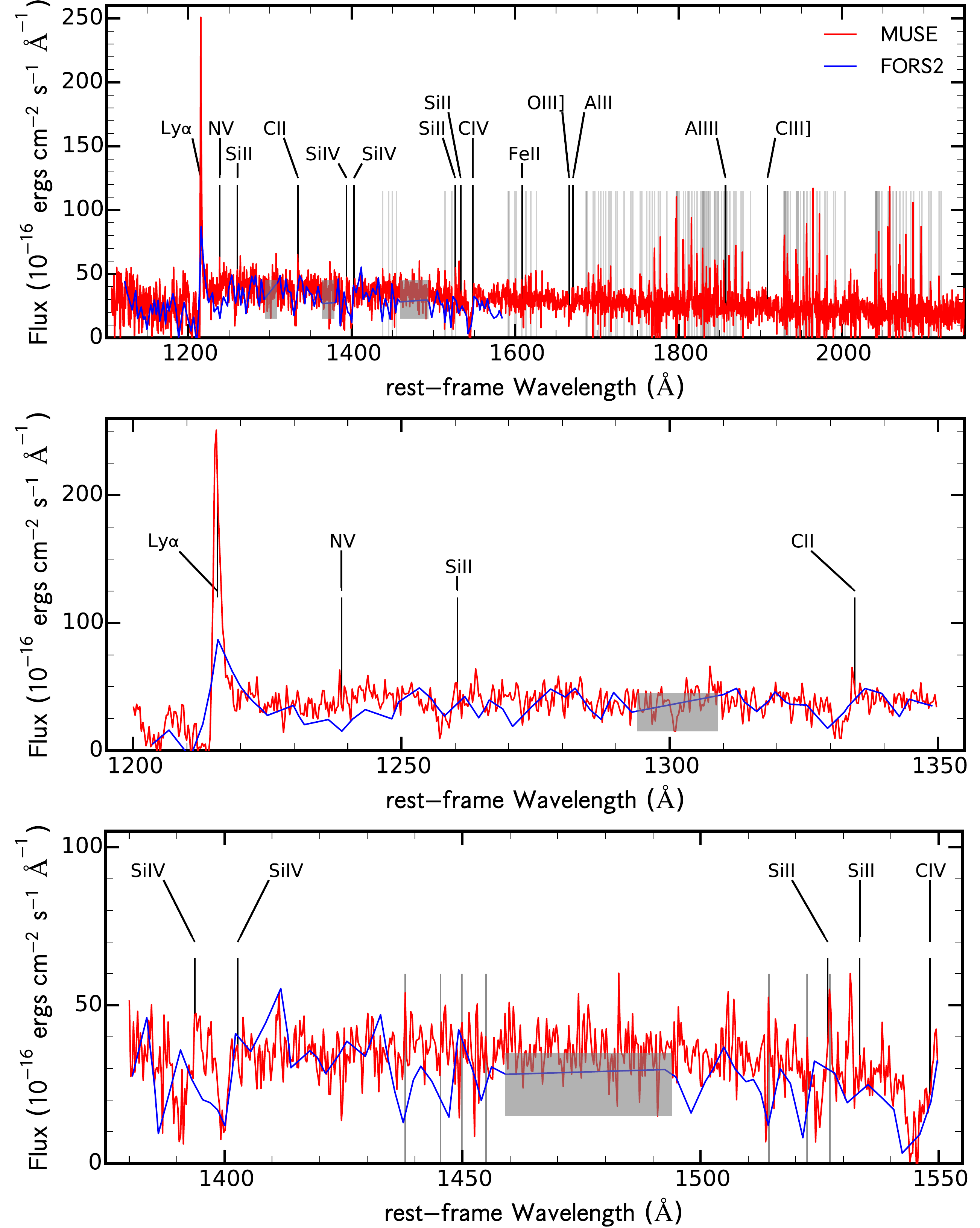}} 
\caption{Comparison between  the MUSE (red) and FORS2 spectra of  \citealt{Iwata+2005}  (blue) for the galaxy \id{43}, a strong \lya\ emitter at $z_{MUSE}=3.2925$. The strongest spectral features are indicated in black. The grey lines show the position of the sky lines; grey areas show  wavelength regions for which no FORS2 spectrum was published. The upper panel shows the entire spectra;   the middle and lower panels show a zoom on the  \lya\ and 1380-1550$\AA$  region, respectively; the latter contains strong  SiIV absorption features.} 
\label{fig:spec_ID_43}
\end{figure}

For two of the galaxies in common with other studies, \id{13} (HDFS J223252.16-603323.9) and  \id{43} (HDFS J223252.03-603342.6), the actual spectra have been published together with the estimated redshifts. This enables us to perform a detailed comparison between these published spectra and our MUSE spectra (integrated over the entire galaxy). This comparison is shown in  Figures~\ref{fig:spec_ID_13} and \ref{fig:spec_ID_43} for \id{13} \citep{Rigopoulou+2005} and \id{43} \citep{Iwata+2005}, respectively.  

The galaxy \id{13} is a strong \oii{3727} emitter at  $z_{MUSE}=$1.2902. In the top panel of Figure~\ref{fig:spec_ID_13}  its full FORS1 and MUSE spectra  are shown in blue and red, respectively. The FORS1 spectrum covers only the rest-frame $\lambda<3500~\mathrm{\AA}$ wavelength region, which does not include the \oii\ feature. The lower panel of the figure shows a zoom  on the rest-frame $\sim$2300-2800$\AA$ window, which includes several strong Fe and Mg absorption features. The MUSE spectrum resolves the FeII and MgII doublets well; it is also clear that some strong features in the FORS1 spectrum, e.g., at  $\sim$2465$\AA$, $2510\AA$, $2750\AA$ and $2780\AA$ are not seen in the high signal-to-noise MUSE spectrum. 

The galaxy  \id{43} is a \lya\ emitter at  $z_{MUSE}=$3.2925. 
 Figure~\ref{fig:spec_ID_43} shows the MUSE spectrum in red and the FORS2 spectrum of \citet{Iwata+2005} in blue.
The middle and bottom panels show zooms on the \lya\ emission line and the rest-frame $\sim$1380-1550$\AA$ Si absorption features,  respectively. The higher SNR and resolution of the MUSE spectrum opens the way to quantitative astrophysical studies of this galaxy, and generally of the early phases of galaxy evolution that precede the $z\sim2$ peak in the cosmic star-formation history of the Universe.

\subsection{Comparison between MUSE and published photometric redshifts}
\label{subsec:photoz}

\begin{figure}
\resizebox{\hsize}{!}{\includegraphics{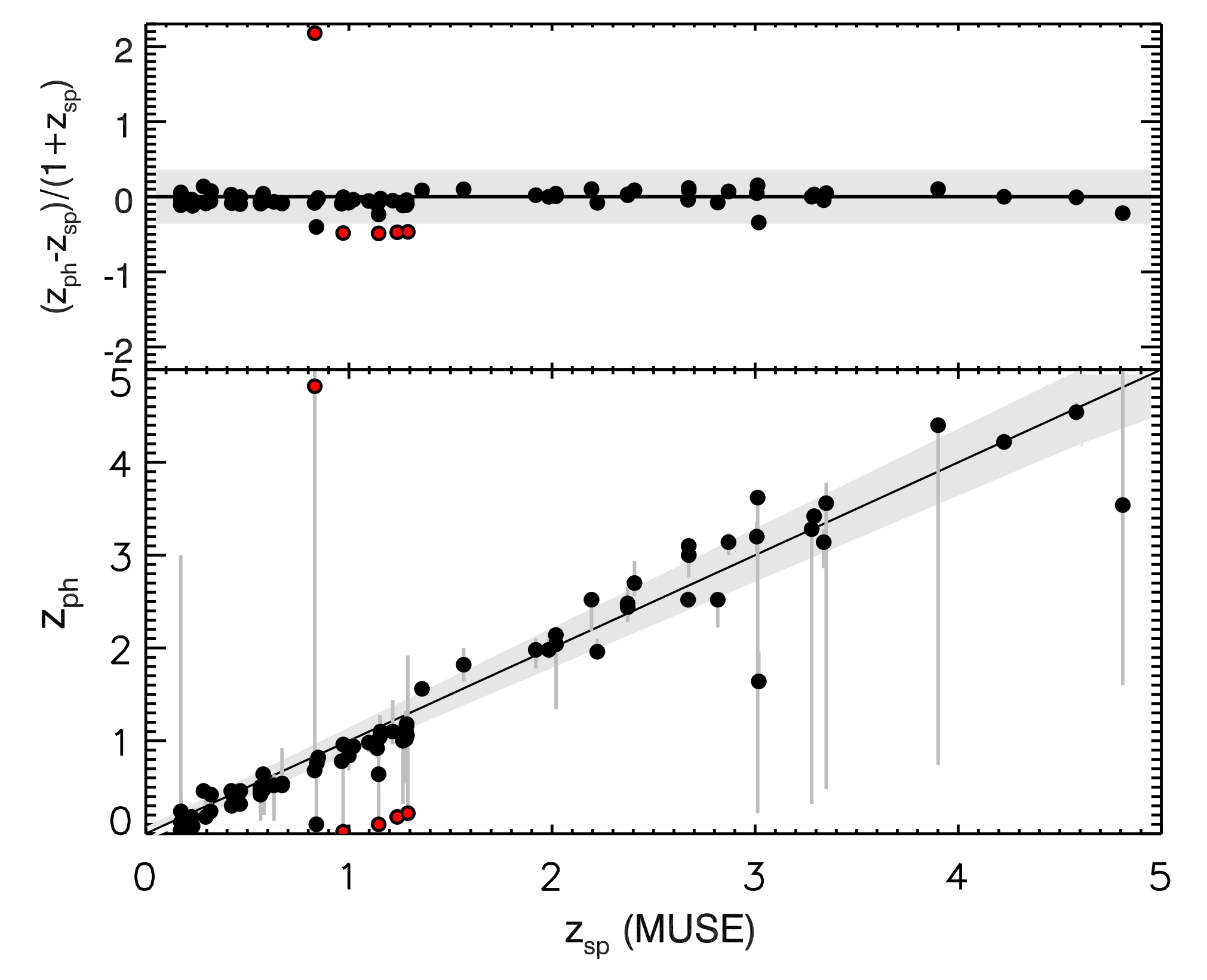}}
\caption[]{Comparison of MUSE spectroscopic redshifts with the
  photometric redshifts of \citet{Labbe+2003}. The upper panel shows
  the distribution of $\Delta$z as a function of MUSE z$_{\rm{{sp}}}$ with
outliers highlighted in red. The error bars shows the uncertainties reported by \citeauthor{Labbe+2003} 
The grey shaded area depicts the region
outside of which objects are considered outliers. The lower panel shows a direct
comparison of MUSE z$_{\rm{{sp}}}$  and \citet{Labbe+2003} z$_{\rm{{ph}}}$ with outliers again highlighted in red.}
\label{fig:zph}
\end{figure}

Our analysis of the HDFS allows for a quantitative comparison with photometric redshifts from the
literature. We make a first comparison to the
photometric redshift catalogue of \cite{Labbe+2003} who used the FIRES
survey to complement existing HST imaging with J, H, and K$_s$ band
data reaching K$_{s, AB}^{tot} \leq 26$. We find $89$ objects in common
between the two catalogues, including $8$ stars. The comparison is given in Figure \ref{fig:zph}.
Considering the $81$
non-stellar objects, we quantify the
agreement between the MUSE spectroscopic redshifts and the 7-band photometric redshifts of
\cite{Labbe+2003} by calculating $\sigma_{\rm{NMAD}}$ (Equation
{\ref{eq:sigNMAD}} of \citealt{Brammer+2008}). This gives the
median absolute deviation of $\Delta z$, and quantifies the number of
`catastrophic outliers', defined as those objects with \mbox{$|\Delta z| >
5\sigma$$_{\rm{NMAD}}$}.

\begin{equation}
{\sigma_{\rm{NMAD}} = 1.48 \times {\rm{median}} \left(\left|\frac{\Delta z-{\rm{median}}(\Delta z)}{1 +  z_{\rm{sp}}}\right|\right)}
\label{eq:sigNMAD}
\end{equation}

\noindent where $\Delta z = (z_{\rm{{sp}}} - z_{\rm{ph}})$.\\

We find \mbox{$\sigma$$_{\rm{NMAD}}= 0.072$} with $6$ catastrophic outliers,
equating to $7.4$
percent of the sample. 
Excluding outliers and recomputing results in $\sigma$$_{\rm{NMAD}}= 0.064$
 reduces the number of
catastrophic outliers to $2$ objects, $2.7$ percent of the remaining
$75$ sources. 

Of the $6$ outlying objects, $5$ are robustly identified as [O{\sc~ii}]
emitters in our catalogue, but the photometric redshifts put all $5$
of these objects at very low redshift, most likely due to template
mismatch in the SED fitting. The final object's spectroscopic redshift
is $z=0.83$, identified through absorption features,
but the photometric redshift places this object at $z=4.82$. This
is not a concern, as the object exhibits a large asymmetric error on
the photometric redshift bringing it into agreement with
z$_{\rm{{sp}}}$. As noted in \citet{Labbe+2003}, this
often indicates a secondary solution to the SED-fitting with
comparable probability to the primary solution, at a very
different redshift.

\begin{figure}
\resizebox{\hsize}{!}{\includegraphics{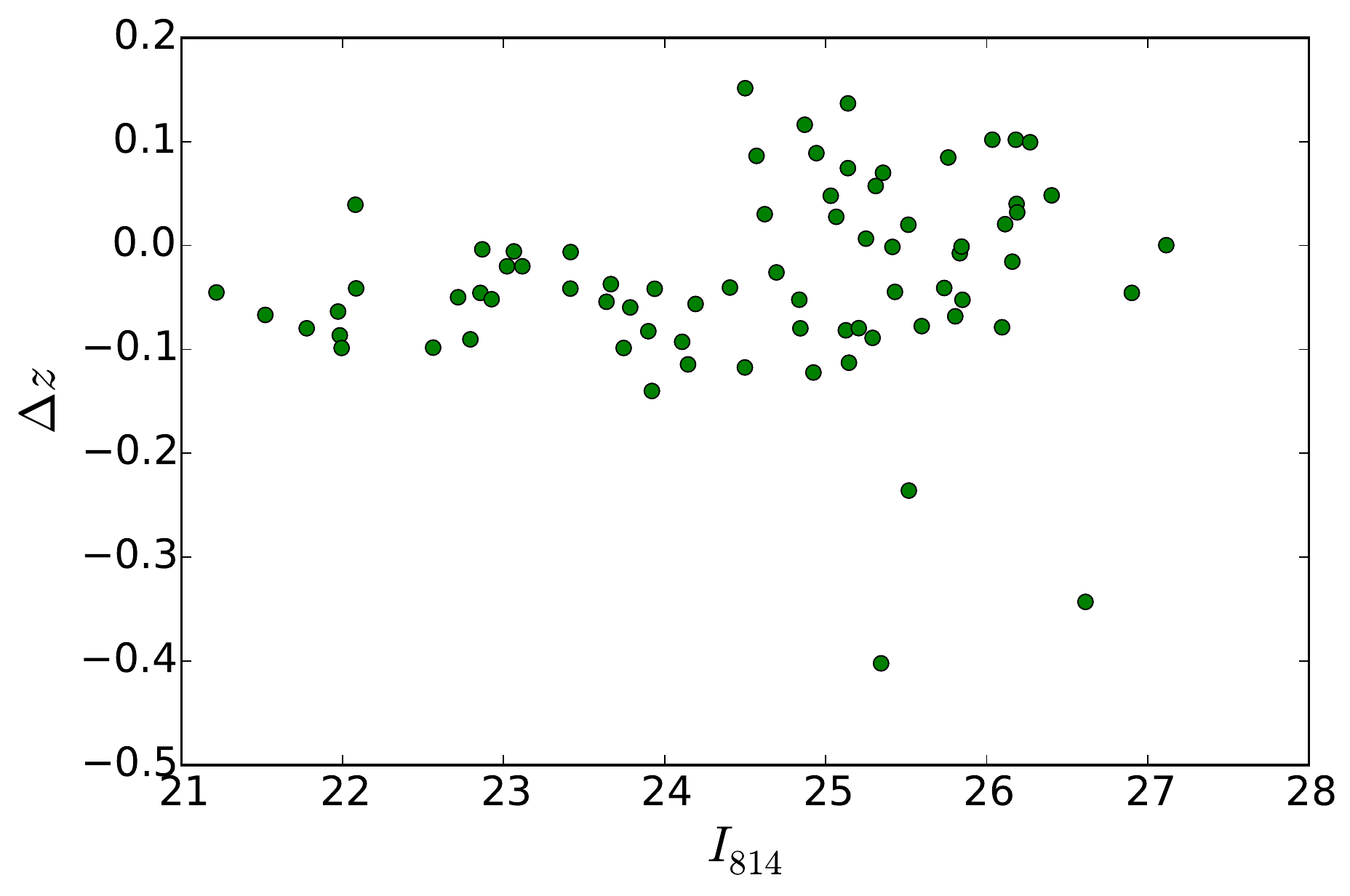}}
\caption[]{Relation between HST \imag\ magnitude and the scatter in
  $\Delta z = (z_{phot}-z_{spec}) / (1 + z_{spec})$. An increase in scatter is seen towards fainter \imag\
  magnitudes, highlighting the importance of spectroscopic redshifts
  for emission-line objects with very faint continuum magnitudes.}
\label{fig:m814}
\end{figure}

The advantage of a blind spectroscopic survey such as ours is
highlighted when considering the reliability of photometric redshifts
for the faint emission-line objects we detect in
abundance here. Figure \ref{fig:m814} shows values of $\Delta z/(1+z)$ for
objects in our catalogue with an HST detection in the F814W
filter. 
For galaxies with magnitude below \imag = 24, the measured scatter (rms) 3.7\%, is comparable to what 
is usually measured (e.g. \citealt{Saracco+2005} and \citealt{Chen+1998}). However, at fainter magnitudes we see an increase with a measured scatter of 11\% (rms) for galaxies
in the 24-27 \imag\ magnitude range, making the photometric redshift less reliable and demonstrating the importance of getting spectroscopic redshifts for faint sources.

\begin{figure}
\includegraphics[width=\linewidth]{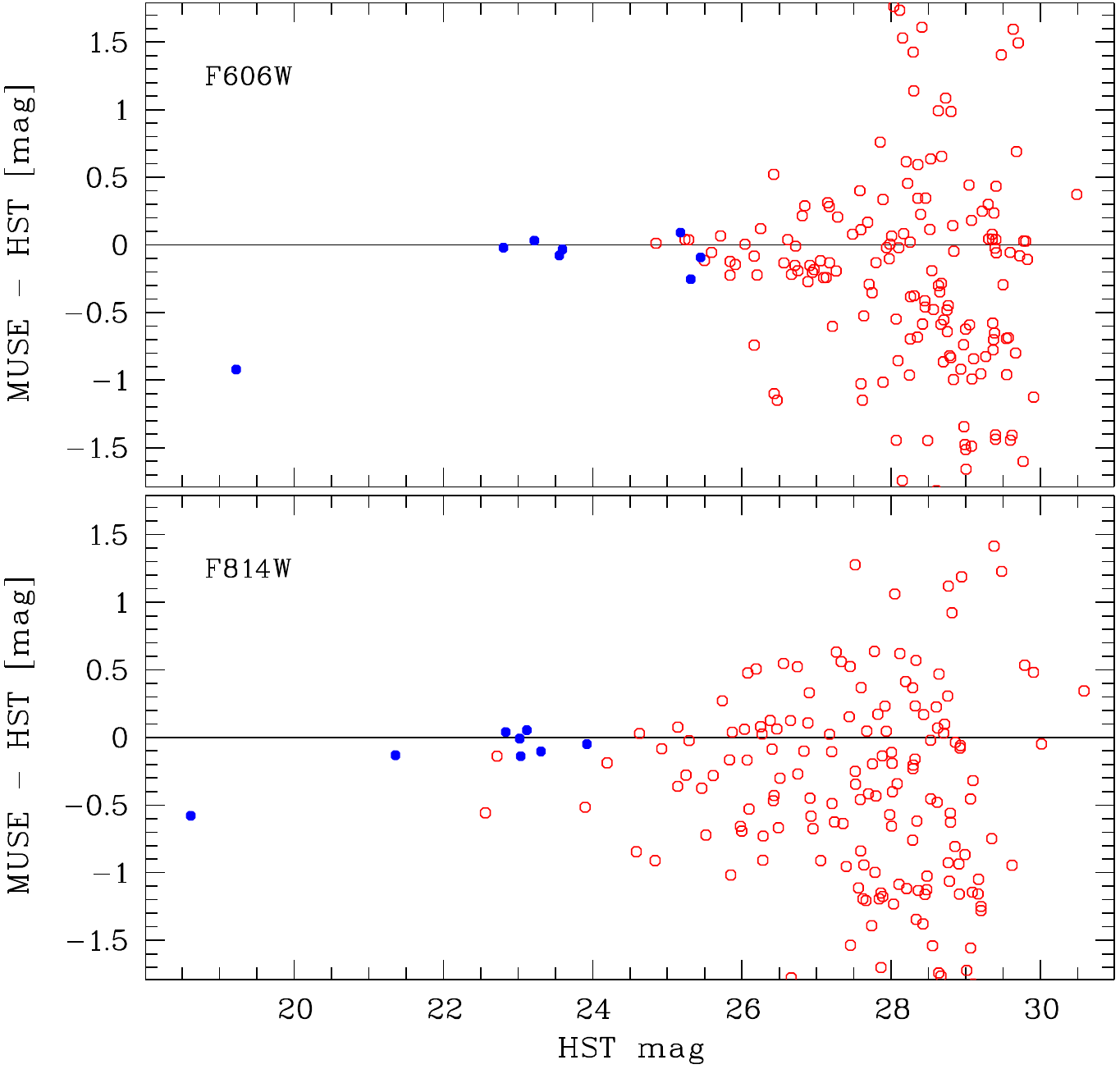}
\caption{Differences between broad-band magnitudes synthesized from extracted MUSE spectra and filter magnitudes measured by HST; top: \rmag, bottom: \imag. The different symbols represent different object types: blue filled circle denote stars and red open circle stand for galaxies with FWHM $<$ 0\farcs4 (in the HST images).} 
\label{fig:lw_bbmag}
\end{figure}

\subsection{Spectrophotometric accuracy}
\label{subsec:lw_specphot}

The spectral range of MUSE coincides almost perfectly with the union of the two HST/WFPC2 filters F606W and F814W. It is therefore possible to synthesize broad-band magnitudes in these two bands directly from the extracted spectra, without any extrapolation or colour terms. In principle, a comparison between synthetic MUSE magnitudes with those measured in the HST images, as provided by \citet{Casertano+2000}, should give a straightforward check of the overall fidelity of the spectrophotometric calibration. In practice such a comparison is complicated by the non-negligible degree of blending and other aperture effects, especially for extended sources but also for the several cases of multiple HST objects falling into one MUSE seeing disk. We have therefore restricted the comparison to stars and compact galaxies with spatial FWHM $<$ 0\farcs4 \citep[as listed by][]{Casertano+2000}. We also remeasured the photometry in the HST data after convolving the images to MUSE resolution, to consistently account for object crowding.

The outcome of this comparison is shown in Fig.~\ref{fig:lw_bbmag}, for both filter bands. Considering only the 8 spectroscopically confirmed stars in the field (blue dots in Fig.~\ref{fig:lw_bbmag}), all of which are relatively bright and isolated, Star \id{0} is partly saturated in the HST images and consequently appears 1~mag brighter in MUSE than in HST. For the remaining 7 stars, the mean magnitude differences (MUSE $-$ HST) are $+$0.05~mag in both bands, with a formal statistical uncertainty of $\pm 0.04$~mag. The compact galaxies (red dots in Fig.~\ref{fig:lw_bbmag}) are much fainter on average, and the MUSE measurement error at magnitudes around 28 or fainter is probably dominated by flat fielding and background subtraction uncertainties. Nevertheless, the overall flux scales are again  consistent. We conclude that the spectrophotometric calibration provides a flux scale for the MUSE datacube that is fully consistent with external space based photometry, without any corrective action.

\section{Census of the MUSE HDFS field}
\label{sect:census}

Given the data volume, its 3D information content and the number of objects found, it would be prohibitive to show all sources in this paper. Instead, detailed informations content for all objects will be made public as described in Appendix~\ref{sect:public}. In this section we carry out a first census of the MUSE data cube  with a few illustrations on a limited number of representative objects.

\begin{figure*}
\centering
 \resizebox{\hsize}{!}{\includegraphics{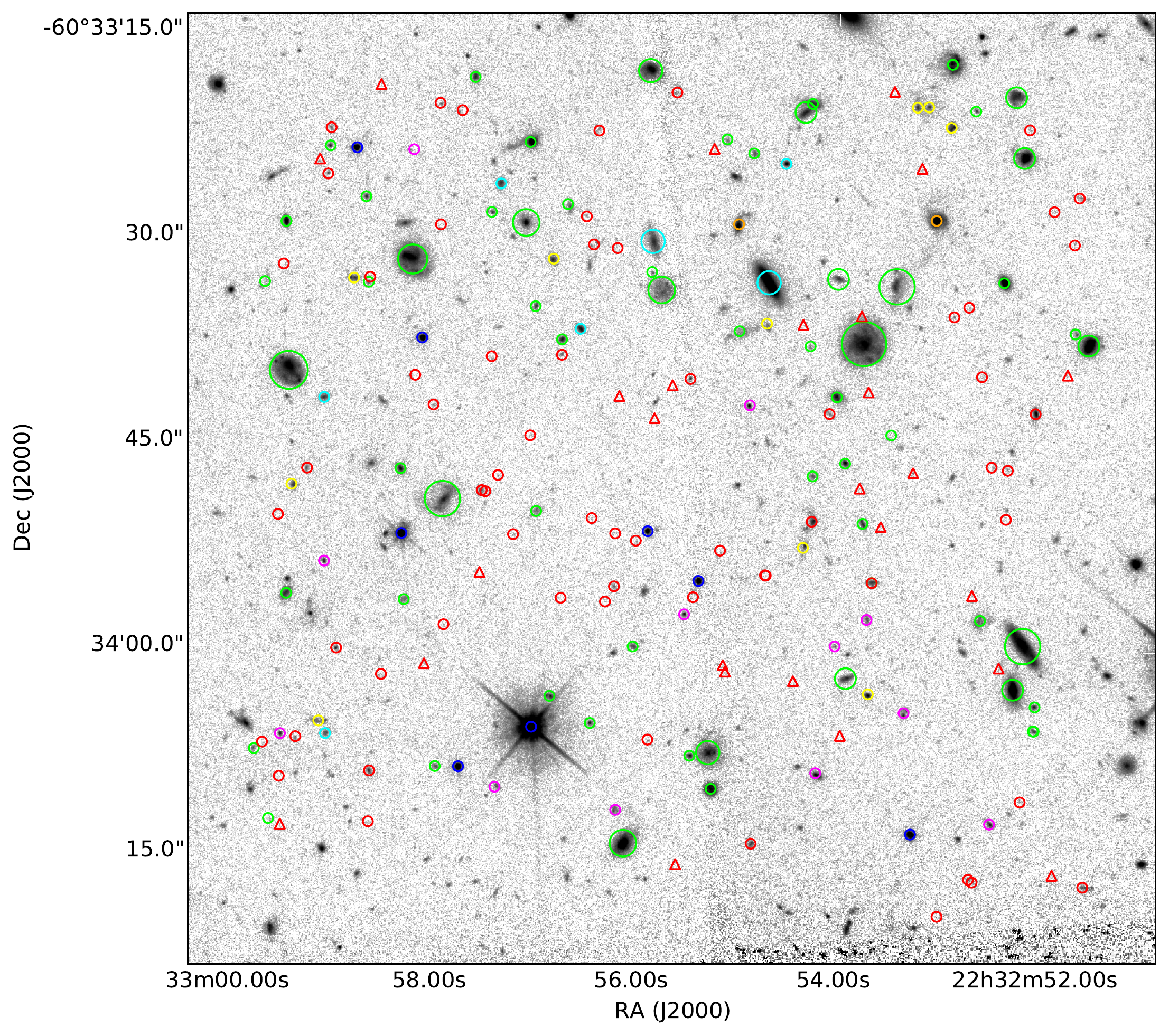}}
\caption{The location of sources with secure redshifts in the HDFS MUSE field. In grey the WFPC2 F814W image. The object categories are identified with the following colors and symbols: 
blue: stars, cyan: nearby objects with $z < 0.3$, green: \oii\ emitters, yellow: objects identified solely with absorption lines, magenta: \ciii\ emitters,
orange: AGN, red circles: \lya\ emitters with HST counterpart, red triangles: \lya\ emitters without HST counterpart. 
Objects which are spatially extended in MUSE are represented by a  symbol with a size proportional to the number of spatially resolved elements.
}
\label{fig:catalog}
\end{figure*}

A total of 189 objects  in the data cube have a securely determined redshift. It is a rich content with 8 stars and 181 galaxies of various categories. Table~\ref{tbl:catalog} and Figure~\ref{fig:catalog} give a global view of the sources in the field.  The various categories of objects are described in the following sub-sections.

\begin{table}[htdp]
\caption{Census of the objects in the MUSE HDFS field sorted by categories.}
\begin{center}
\begin{tabular}{lrcc}
\hline \hline
Category & Count & z range & \imag\ range\\
\hline
Stars & 8 & 0 & 18.6 - 23.9 \\
Nearby galaxies & 7 & 0.12 - 0.28 & 21.2 - 25.9 \\
$[O\,\textsc{ii}]$ emitters & 61 & 0.29 - 1.48 & 21.5 - 28.5 \\
Absorption lines galaxies & 10 & 0.83 - 3.90 & 24.9 - 26.2 \\
AGN & 2 & 1.28 & 22.6 - 23.6 \\
$C\,\textsc{iii}]$ emitters & 12 & 1.57 - 2.67 & 24.6 - 27.2 \\
\lya\ emitters & 89 & 2.96 - 6.28 & 24.5 - 30+ \\
\hline
\end{tabular}
\end{center}
\label{tbl:catalog}
\end{table}%

\subsection{Stars}
We obtained spectra for 8 stars in our field. Seven were previously identified by \citet{Kilic+2005} from proper motion measurements of point sources in the HDFS field. Among these stars we confirm that HDFS 1444 (\id{18}) is a white dwarf. We also identify one additional M star (\id{31}) that was not identified by \citet{Kilic+2005}.

\subsection{Nearby galaxies}
In the following  we refer to  objects whose 
\oii\ emission line is redshifted below the 4800 \AA\ blue cut-off of
MUSE as nearby galaxies; that is all galaxies with $z < 0.29$. Only 7 galaxies fall into
this category. Except for a bright (\id{1} \rmag = 21.7) and a
fainter (\id{26} \rmag = 24.1) edge-on disk galaxy,  the 5  remaining objects are faint compact dwarfs (\rmag $\sim 25-26$).

\subsection{\oii{} emitters}

\begin{figure*}
\centering
 \resizebox{\hsize}{!}{\includegraphics{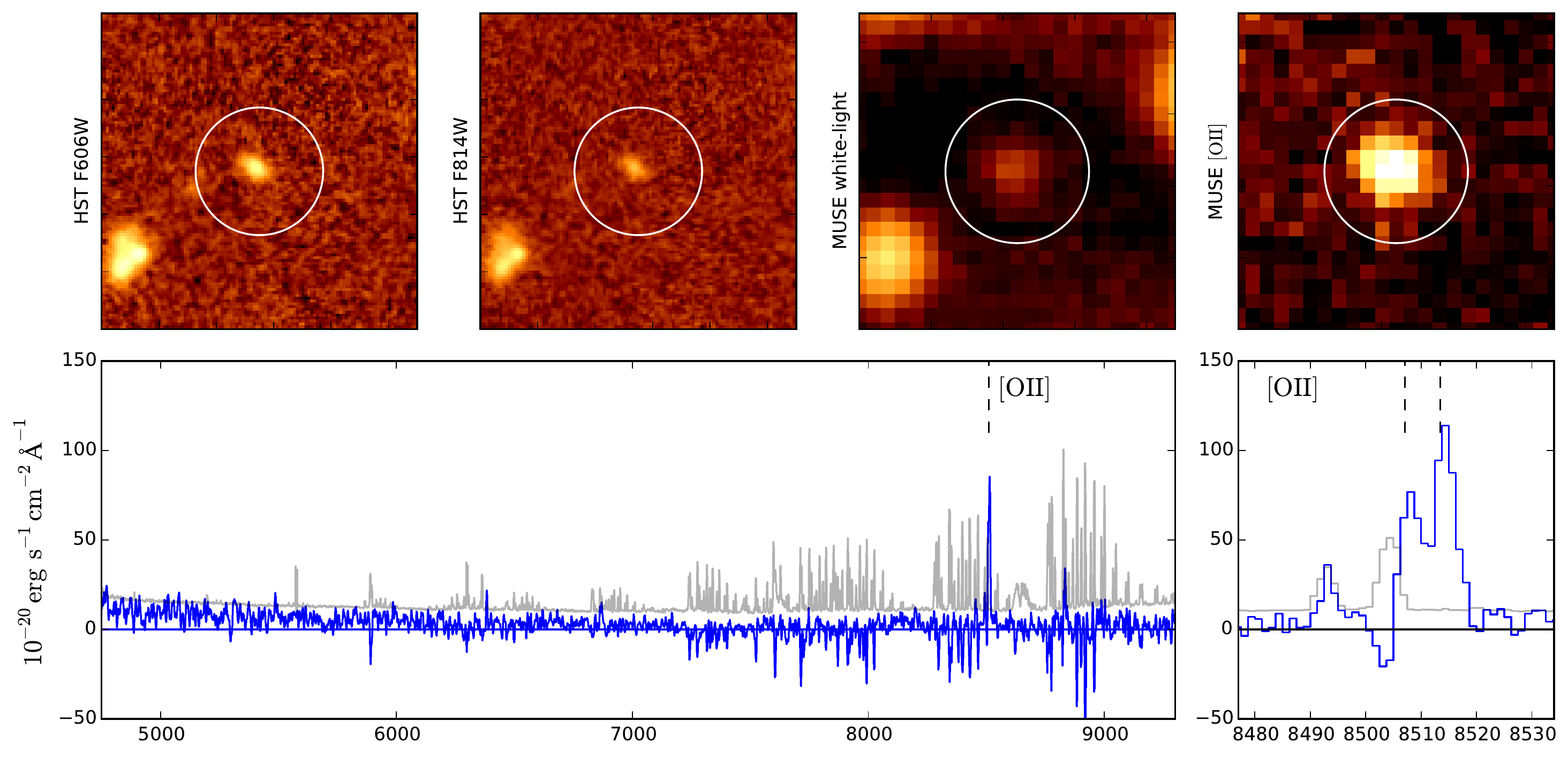}}
\caption{\id{160} is a  $z=1.28$  \oii\  emitter with a faint (\imag $\sim 26.7$)  HST counterpart. The HST images in the F606W and F814W filters are shown at the top left, the  MUSE reconstructed white-light and \oii\  continuum subtracted narrow band images at the top right. The one arcsec radius red circles show the object location derived from the HST image. 
At the bottom left, the full spectrum (in blue), smoothed with a 4\AA\ boxcar, and its 3$\sigma$ error (in grey) are displayed.  A zoom of the unsmoothed spectrum, centered around the \oii{3726,3729} emission lines, is also shown at the bottom right.
}
\label{fig:example-OII}
\end{figure*}

A large fraction of identified galaxies have \oii{3726,3729} in emission and we will refer to these as \oii{}-emitters in the following even if this is not the strongest line in the spectrum. In Fig.~\ref{fig:example-OII} we show an example of a faint \oii\ emitter at z=1.28 (\id{160}). In the HST image it is a compact source with a 26.7 \rmag\ and \imag\ magnitude. 

The average equivalent width of \oii{3727} is 40\AA\ in galaxies spanning a wide range in luminosity from dwarfs with $M_B\approx -14$ to the brightest galaxy at $M_B\approx -21.4$, and sizes from marginally resolved to the largest (\id{4}) with an extent of 0.9" in the HST image. 

It is also noticeable that the \oii{}-emitters often show significant Balmer absorption. In the D4000$_N$--H$\delta_A$ diagram they fall in the region of star forming and post-starburst galaxies. This frequent strong Balmer absorption does fit with previous results from the GDDS \citep{LeBorgne+2007} and VVDS \citep{Wild+2009} surveys.

\subsection{Absorption line galaxies}

For 10 galaxies, ranging from $z=0.83$ to $z=3.9$, the redshift determination has been done only on the basis of absorption lines. This can be rather challenging for faint sources because establishing the reality of an absorption feature is more difficult than for an emission line.  For that reason the faintest source with a secure absorption line redshift has \imag$=26.2$ and $z=3.9$, while the faintest source (\id{83}) with absorption redshifts in the $1.5<z<2.9$  so-called MUSE  `redshift desert' \citep{Steidel+2004} has \imag\ of 25.6. 

A notable pair of objects is \id{50} and \id{55} which is a merger at $z=2.67$ with  a possible third companion based on the HST image, which can not be separated in the MUSE data. And while not a pure absorption line galaxy, as it does have \oii{3727} in emission, object \id{13} shows very strong Mg$\,$\textsc{ii}$\, and\, Fe$\,$\textsc{ii}$ absorption lines.

\subsection{\ciii\ emitters}

\begin{figure*}
\centering
 \resizebox{\hsize}{!}{\includegraphics{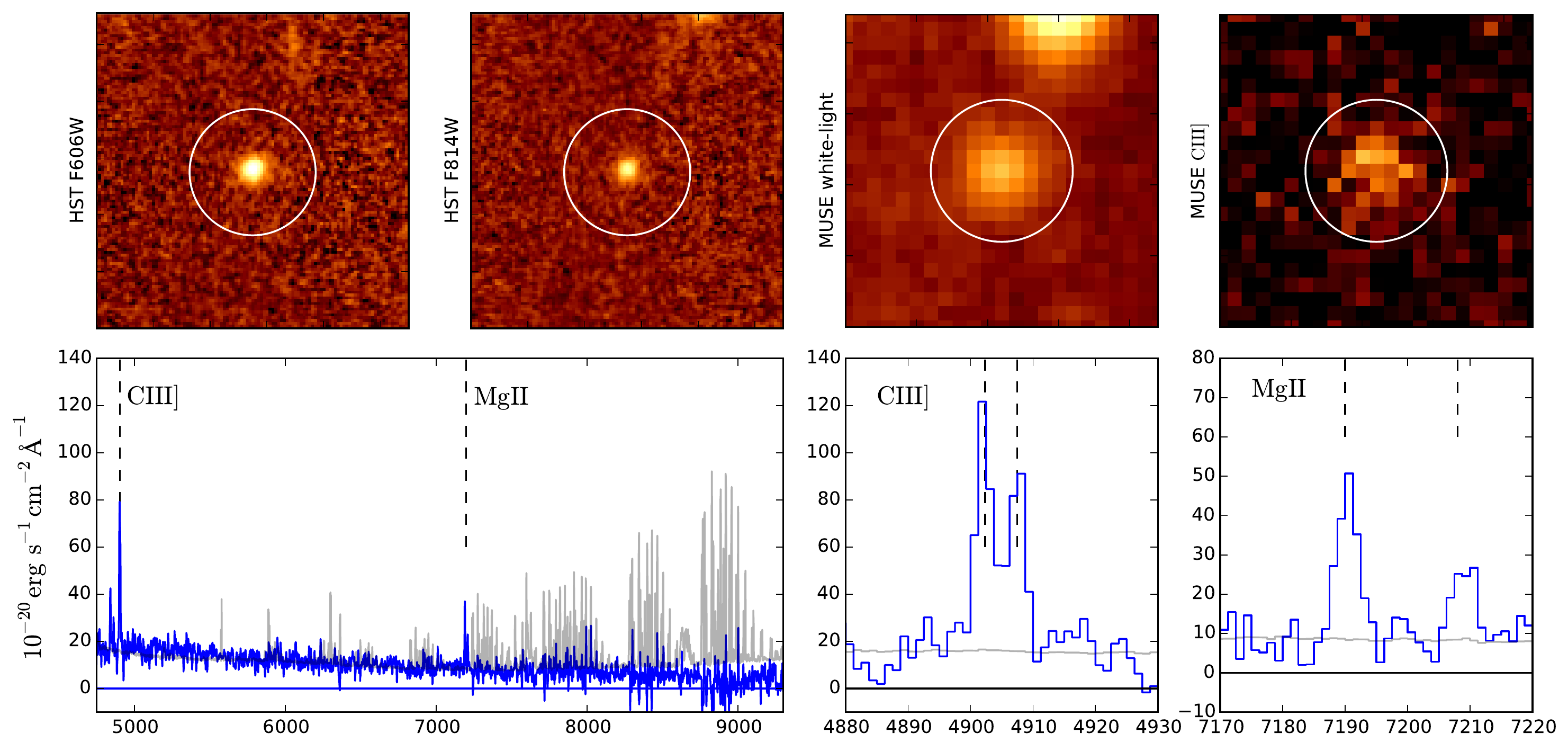}}
\caption{\id{97} is a z=1.57 strong \ciii\  emitter. The HST images in F606W and F814W filters are shown at the top left, the  MUSE reconstructed white-light, the \ciii\  and \mgii\ continuum subtracted narrow band images at the top right. The one arcsec radius red circles show the object location derived from the HST image.
At the bottom left, the full spectrum (in blue), smoothed with a 4\AA\ boxcar, and its 3$\sigma$ error (in grey) are displayed.  A zoom of the unsmoothed spectrum, centered around the \ciii{1907,1909} \AA\  and  \mgii{2796,2803} \AA\ emission lines, are also shown at the bottom right.}
\label{fig:example-CIII}
\end{figure*}

At $1.5<z<3$, well into the  `redshift desert',  the main emission line identified is \ciii{1907,1909}, which is typically resolved as a doublet 
of emission lines at the resolution achieved with MUSE. Among the clear \ciii\ emitters identified, the most interesting one (\id{97}) is displayed in Fig.~\ref{fig:example-CIII}. 
It is a z=1.57 galaxy with strong \ciii{1907,1909}  and \mgii{2796,2803} emission lines. It appears as a compact source in the HST images with \rmag$= 26.6$ 
and \imag$= 25.8$. The object is unusually bright in \ciii\ with a total flux of  $2.7 \times 10^{-18} erg.s^{-1}.cm^{-2}$ and a rest-frame equivalent width of 16 \AA. These are relatively 
rare objects,  with only 17 found in our field of view, but such \ciii\ emission is expected to appear for younger and lower mass galaxies, typically showing a high ionization parameter 
\citep{Stark+2014}.

\subsection{\lya\ emitters}
 \label{sect:stackedlya}
\begin{figure*}
\centering
 \resizebox{\hsize}{!}{\includegraphics{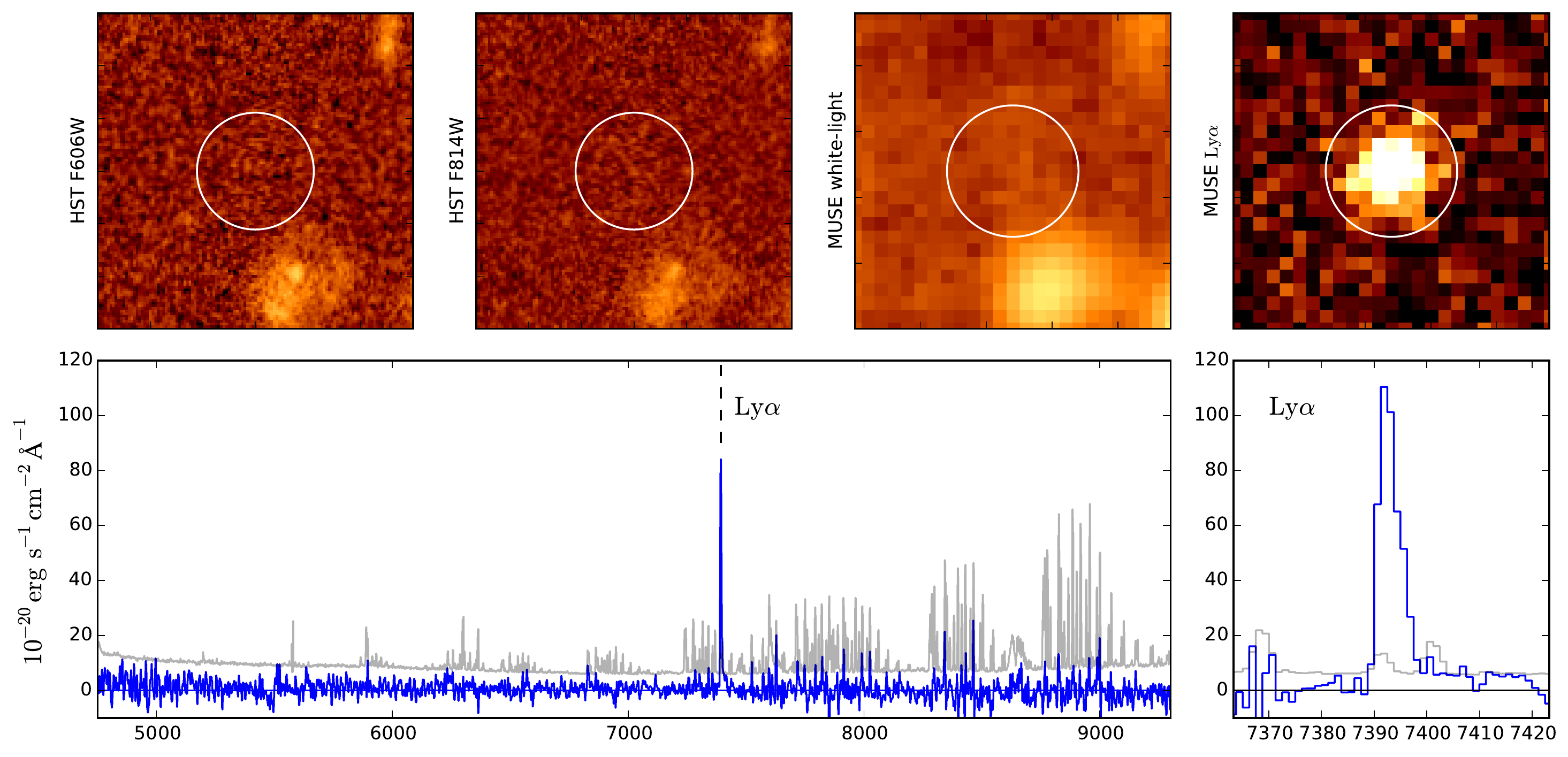}}
\caption{\id{553} is a z=5.08 \lya\  emitter without HST counterpart. The HST images in F606W and F814W filters are shown at the top left, the  MUSE reconstructed white-light and \lya\ narrow band images at the top right. The one arcsec radius red circles show the emission line location. The spectrum is displayed on the bottom figures; including a zoom at the emission line.
At the bottom left, the full spectrum (in blue), smoothed with a 4\AA\ boxcar, and its 3$\sigma$ error (in grey) are displayed.  A zoom of the unsmoothed spectrum, centered around the \lya\ emission line, is also shown at the bottom right.
}
\label{fig:example-lya}
\end{figure*}

The large majority of sources at $z>3$ are identified through their strong Lyman-$\alpha$ emission line. Interestingly, 
26 of the discovered \lya\ emitters are below the HST detection limit, i.e \rmag$>29.6$ and \imag$>29$ (3$\sigma$ depth in a 0.2 $arcsec^2$ aperture, \citealt{Casertano+2000}).  We produced a stacked image in the WFPC2-F814W filter of these 26 \lya\ emitters not individually detected in HST and measured an average continuum at the level of \imag$ = 29.8 \pm 0.2$ AB (Drake et al. in prep.). 
We present in Fig.~\ref{fig:example-lya} one such example, \id{553} in the catalog. Note the typical asymmetric \lya\ profile. With a total \lya\ flux of $4.2 \times 10^{-18} erg\,s^{-1}cm^{-2}$ the object is one of the brightest of its category. It is also unambiguously detected in the reconstructed \lya\ narrow band image. With such a low continuum flux the 
emission corresponds to a rest-frame equivalent width higher than $130\, \AA$.

Note that we have found several even fainter line emitters that have no HST counterpart. However, because of their 
low SNR, it is difficult to firmly identify the emission line and they have therefore  been discarded from the final catalog.

\subsection{Active Galactic Nuclei}
\label{subsec:lw_agn}

Among the \oii\ emitting galaxies we identify two objects (\id{10} and \id{25}) that show significant [\ion{Ne}{V}] 3426 emission, a strong signature of nuclear activity. Both galaxies show pronounced Balmer breaks and post-starburst characteristics, and their forbidden emission lines are relatively broad with a FWHM $\sim$230 \kms. There is, however, no clear evidence for broader permitted lines such as \mgii{2798}, thus both objects are probably type~2 AGN. Both objects belong to the same group of galaxies at $z\simeq 1.284$ (Sect.\ref{sect:groups}). 

Object \id{144} was classified by \citet{Kilic+2005} as a probable QSO at $z=4.0$ on the basis of its stellar appearance and its \emph{UBVI} broad band colours. The very strong \lya\ line of this objects confirms the redshift ($z = 4.017$), but as the line is relatively narrow ($\sim$100 \kms) and no other typical QSO emission lines are detected, a definite spectroscopic classification as an AGN is not possible.

\subsection{Spatially resolved galaxies} 
Twenty spatially resolved galaxies up to $z \sim 1.3$ are identified in the MUSE data cube (see Fig.~\ref{fig:catalog}). 
We consider a galaxy as  resolved if it extends over a minimum area of twice the PSF. 
To compute this area we performeded an emission line fitting (see Sect.\ref{sect:resolved}) over a list of 33 galaxies 
that had previously been identified  to be extended in the HST images. Flux maps were built for each fitted emission 
line and we computed the galaxy size (FWHM of a 2D fitted gaussian) using the brightest one (usually \oii).  
 Among these 20 resolved galaxies, 3 are at low redshift ($z\leq 0.3$) and 4 are above $z\sim 1$. 
 Note that 5 of the resolved galaxies are in the group identified at $\sim 0.56$  (see sect.\ref{sect:groups}) including 
 \id{3} which extends over $\sim 5$ times the MUSE PSF. 
 
\subsection{Overlapping objects}
\begin{figure*}
\centering
 \resizebox{\hsize}{!}{\includegraphics{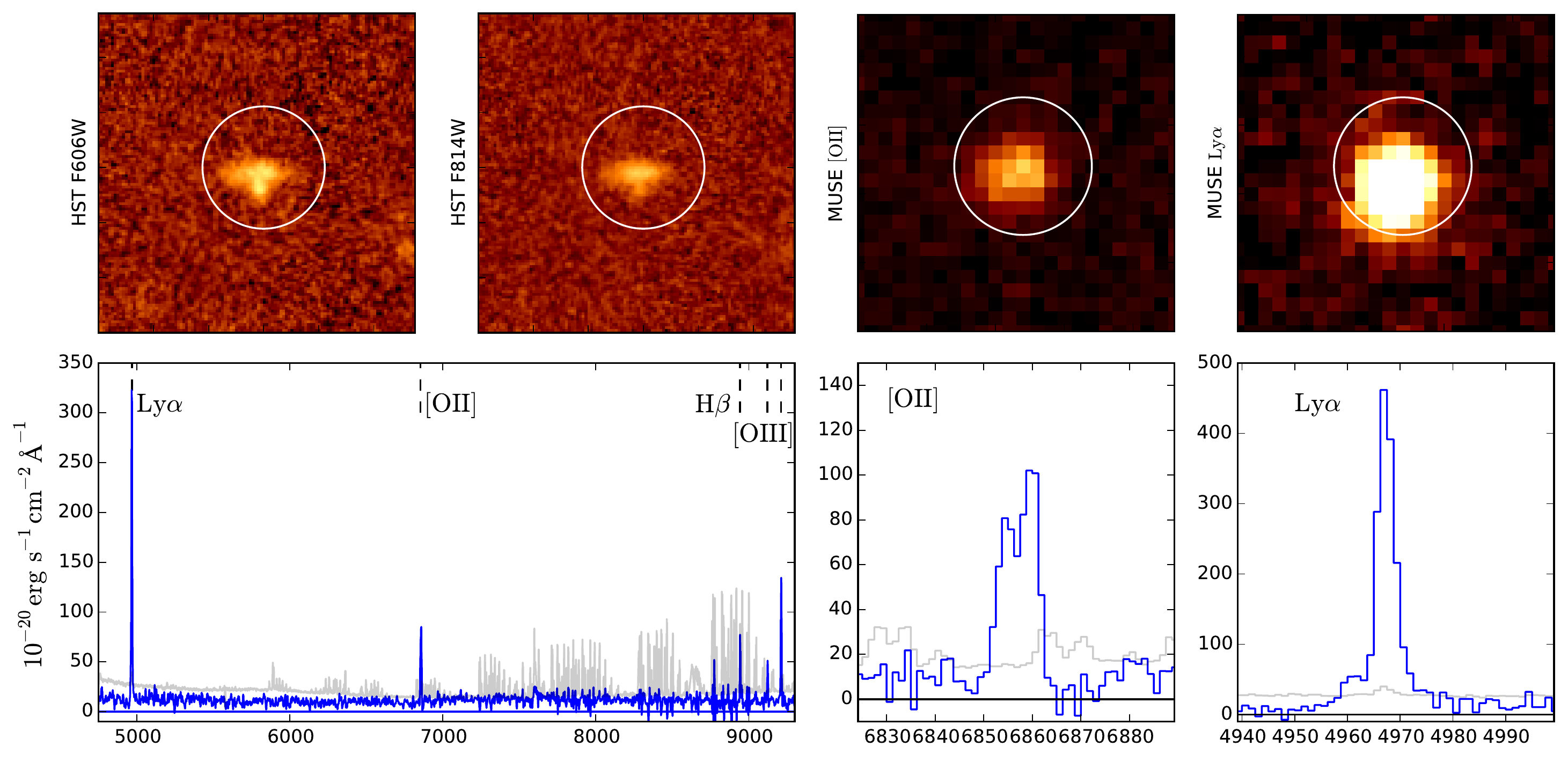}}
\caption{Example of spatially overlapping objects: a z=3.09 \lya\ emitter (\id{71})  with a z=0.83 [OII] emitter (\id{72}). The HST images in F606W and F814W filters are shown at the top left, the MUSE \lya\ and \oii\  images at the top right. The one arcsec radius red circles show the object location derived from the HST image. 
At the bottom left, the full spectrum (in blue), smoothed with a 4\AA\ boxcar, and its 3$\sigma$ error (in grey) are displayed.  A zoom of the unsmoothed spectrum, centered around the \lya\ and \oii{3727} emission line, is also shown at the bottom right.
}
\label{fig:example-Overlap}
\end{figure*}

While searching for sources in the MUSE data cube, we encountered a number of spatially overlapping objects. In many cases a combination of high spatial resolution HST images and MUSE narrow-band images has been sufficient to assign spectral features and redshifts to a specific galaxy in the HST image. But in some cases, the sources cannot be disentangled, even at the HST resolution. This is illustrated in Fig.~\ref{fig:example-Overlap}, where the HST image shows only one object but MUSE reveals it to be the result of two galaxies that are almost perfectly aligned along the line of sight: an \oii{} emitter at $z=0.83$ and a $z=3.09$ \lya\ emitter. There are other cases of objects that potentially could be identified as mergers on the basis of the HST images but are in fact just two galaxies at different redshifts. The power of the 3D information provided by MUSE is nicely demonstrated by these examples.

\section{Redshift distribution and global properties}
\label{sect:redshifts}

\subsection{Redshift distribution}

\begin{figure*}
\centering
 \resizebox{\hsize}{!}{\includegraphics{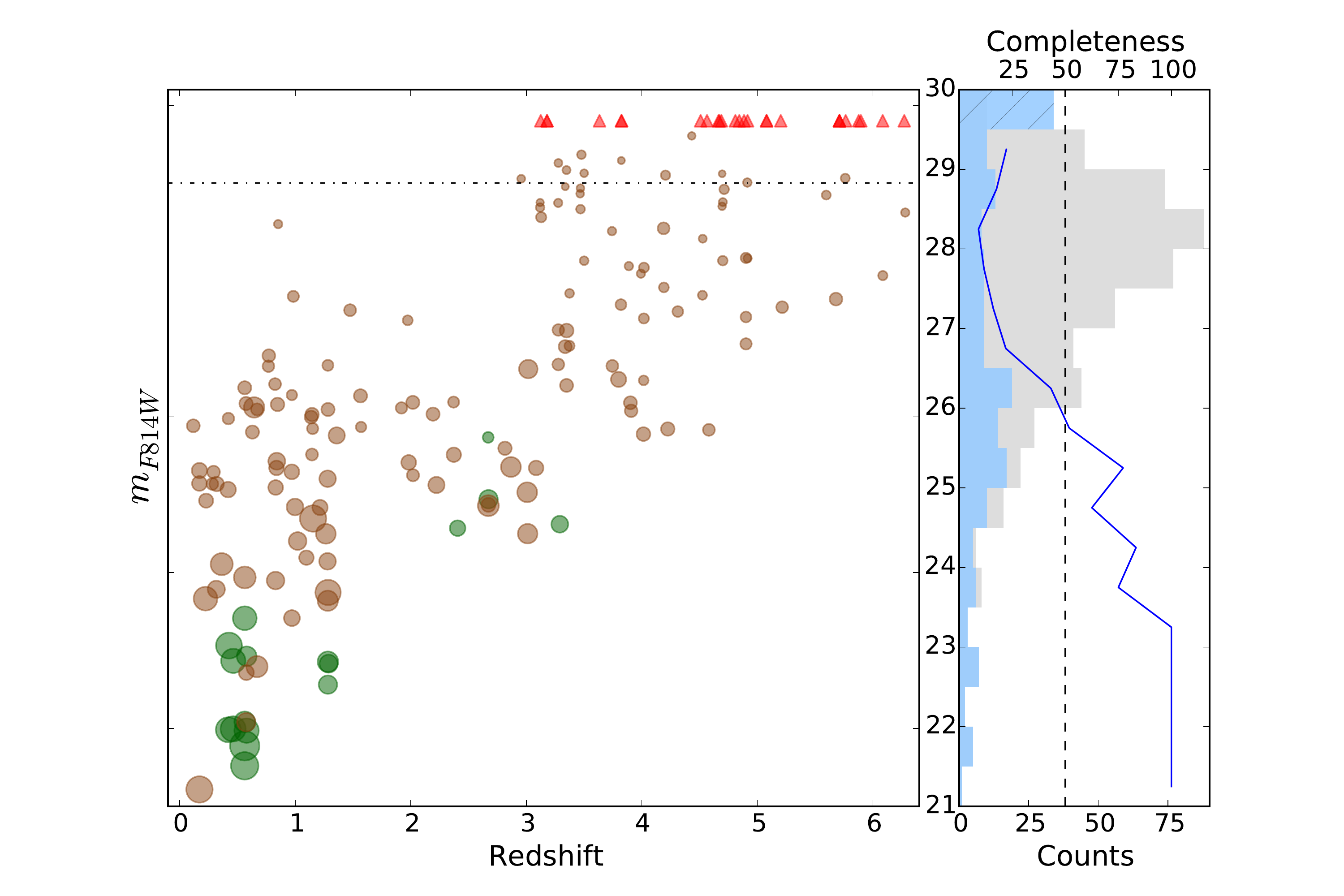}}
\caption{Redshift distribution of sources in the HDFS MUSE field. Left: redshift versus HST \imag AB magnitude. The symbol size is proportional to the object's HST size. The green circles are for the published spectroscopic redshifts and the brown symbols for the new MUSE measurements.
The red arrows show objects without HST counterparts. The dashed line shows the $3\sigma$ detection limit for the HST F814W images.
Right: the gray histogram shows the distribution of all magnitudes in the HST catalogue while the light blue histogram shows the magnitude distribution of those galaxies with confirmed redshifts and HST counterparts. The last histogram bar centered at \imag$=29.75$  is for the \lya\ emitters not detected in the HST image. The blue curve gives the completeness of the redshift identification with respect to the identified sources in the HST images.
}
\label{fig:redshifts}
\end{figure*}

We have been able to measure a redshift at confidence $\ge 1$  for 28\% of the 586 sources
reported in HST catalog of \citet{Casertano+2000} in the MUSE field.
The redshift
distribution is presented in Fig.~\ref{fig:redshifts}. 
We reach 50\% completeness with respect
to the HST catalog at \imag$= 26$. At fainter magnitudes the completeness decreases, but it
is still around 20\% at \imag$= 28$. In addition to the sources
identified in the HST images, we found 26 \lya\ emitters, i.e. 30\% of
the entire \lya\ emitter sample, that have no HST counterparts and
thus have \imag$ > 29.5$.

Redshifts are distributed over the full $z=0-6.3$ range. Note the decrease in the $z=1.5-2.8$ window -- the well known redshift desert -- corresponding to the wavelengths where \oii\ is beyond the 9300 \AA\ red limit of MUSE and \lya\  is bluer than the 4800 \AA\ blue cut-off of MUSE.  

\begin{figure}
\centering
 \resizebox{\hsize}{!}{\includegraphics{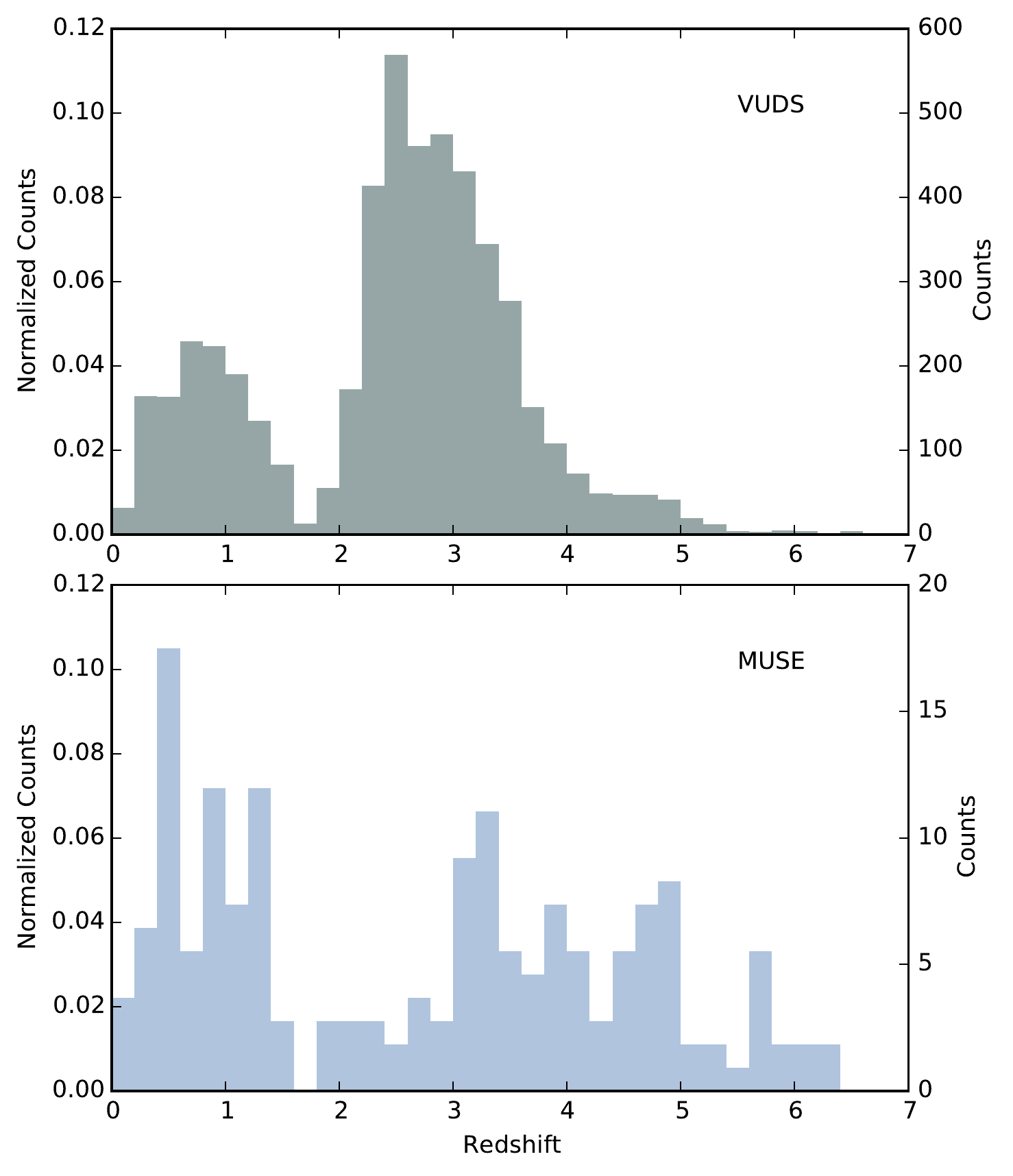}}
\caption{Comparison of the HDFS-MUSE normalized redshift distribution (bottom) with the one from the VUDS (top).
}
\label{fig:compVUDS}
\end{figure}

Although the MUSE HDFS field is only a single pointing, and thus prone to cosmic variance, one can compare the measured redshift distribution with  those from other deep spectroscopic surveys (e.g. zCOSMOS-Deep - \citealt{Lilly+2007}, \citealt{Lilly+2009}, VVDS-Deep - \citealt{Lefevre+2013}, VUDS - \citealt{Lefevre+2014}).  The latter, with 10,000 galaxies in the $z \sim 2 - 6$ range, is the most complete. 
We show in Figure \ref{fig:compVUDS} the MUSE-HDFS and VUDS normalized redshift distributions. 
They look quite different. This was expected given the very different observational strategy: 
VUDS is continuum pre-selected and produces some particular redshift distribution based on the interplay of various factors, while MUSE  does not make any pre-selection.
With 22\% of galaxies at $z > 4$ in contrast to 6\% for the VUDS, MUSE demonstrates a higher efficiency for finding high redshift galaxies. 
The number density of observed galaxies is also very different between the two types of observations. With 6000 galaxies per square degree, the VUDS is the survey which achieved the highest density with respect to the other spectroscopic surveys (see Figure 30 of
\citealt{Lefevre+2014}). This translates, however, to only 1.7 galaxies per $arcmin^2$, which is two orders of magnitude smaller than the number density achieved by MUSE in the HDFS.

\subsection{Galaxy groups}
\label{sect:groups}
The high number density of  accurate spectroscopic redshifts 
allows us to search for galaxy groups in the field.  
We applied a classical friend-of-friend algorithm to identify galaxy
groups, considering only secure redshifts in our catalog,
i.e. those with confidence index $\ge 1$. We adopted a maximum linking
length $\Delta r$ of 500\,kpc in projected distance and $\Delta
v=$700\,km\,s$^{-1}$ in velocity following the zCosmos high redshift
group search by \cite{Diener+2013}. The projected distance
criterium has very little influence since our field-of-view of 1' on
a side corresponds to $\sim$500 kpc at most.  This results in the
detection of 17 groups with more than three members. These are listed
in Table~\ref{table_groups} together with their richness, redshift, and
nominal rms size and velocity dispersion as defined in \citealt{Diener+2013}. 

Among the 181 galaxies with a secure redshift in our catalog, 43\%
reside in a group, 29\% in a pair and 28\% are isolated. The
densest structure we find lies at $z=$1.284 and has nine members,
including 2 AGN and an interacting system showing a tidal tail, all
concentrated to the north west of the field.  
The structure at $z\sim$0.578 
first spotted by \cite{Vanzella+2002}  and further identified as a rich cluster 
by \cite{Glazebrook+2006} shows up here as a 5-member group. The richer
group (6 members) at slightly lower redshift ($z$=0.565) is also within the redshift range 0.56--0.60
considered by \cite{Glazebrook+2006} and could be part of the same large scale structure.
The highest redshift
group identified is also the one with the lowest $v_{rms}$ of all
groups: three Ly-$\alpha$ emitters at z=5.71 within 26
km\,s$^{-1}$. We also find two relatively rich groups both including six
members at redshift $\sim$4.7 and $\sim$4.9.

\begin{table}
  \caption{\label{table_groups} Galaxy groups detected in the HDFS ordered by redshift.}
  \centering
  \begin{tabular}{c r r c l}
\hline
\hline
  $z$ & $v_{rms}$         & $r_{rms}$  & $N_m$ & Member IDs \\
      & km s$^{-1}$& kpc  &&\\
\hline
   0.172&    65& 43 & 3 &       1, 63, 70 \\
   0.421&    262& 54 & 4 &       6, 57, 101, 569\\
   0.564&    52 & 142& 7 &       3, 4, 9, 23, 32, 135\\
   0.578&    424& 150& 5 &       5,  8, 11, 17, 122\\
   0.972&    56 & 201& 3 &      24, 68, 129\\
   1.284&    354& 92 & 9 &      10, 13, 15, 25, 27, 35,\\
   & & & &  64, 114, 160\\
   2.672&    101& 87 & 4 &      50, 51, 55, 87\\
   3.013&    350& 115& 3 &      40, 56, 155\\
   3.124&    329& 92 & 4 &     422, 437, 452, 558\\
   3.278&    36 & 144& 4 &     162, 202, 449, 513\\
   3.349&    35 & 90 & 3 &     139, 200, 503\\
   3.471&    324& 139& 4 &     433, 469, 478, 520\\
   3.823&    161& 93 & 4 &     238, 514, 563, 581\\
   4.017&    113& 181& 4 &      89, 144, 216, 308\\
   4.699&    430& 109& 6 &     325, 441, 453, 474, 499, 548\\
   4.909&    370& 164& 6 &     186, 218, 334, 338, 484, 583\\
   5.710&    26 & 101& 3 &     546, 547, 574\\
\hline
\end{tabular}
\end{table}

\subsection{Number counts of emission line galaxies}
\label{subsec:lw_nc}

With more than 100 emission line galaxies identified by MUSE in this field, a meaningful comparison of the observed number counts with expectations from the literature is  possible. We consider \lya\ and \oii{3727}  emitters.

While integrated emission line fluxes can, in principle, be measured easily from the fits to the extracted spectra (see Appendix \ref{sect:catalog}), the significant level of crowding among faint sources resulted in extraction masks that were often too small for capturing the total fluxes. This is especially relevant for \lya\ emitters which in many cases show evidence for extended line emission. In order to derive robust total fluxes of \oii\ and \lya\ emitters we therefore adopted a more manual approach. We first produced a ``pure emission line cube'' by median-filtering the data in spectral direction and subtracted the filtered cube. We then constructed narrow-band images centered on each line, which were now empty of sources except for the emitters of interest. Finally, total line fluxes were determined by a growth curve analysis in circular apertures around each source.
We adjusted the local sky background for each line such that the growth curve became flat within $\sim$2\arcsec\ from the centroid, and integrated the emission line flux inside that aperture. Avoiding the edges of the field, this procedure yielded subsamples of 74 \lya\ and 41 \oii\ emitters,  with well-measured emission line fluxes.

\begin{figure}
\includegraphics[width=\linewidth]{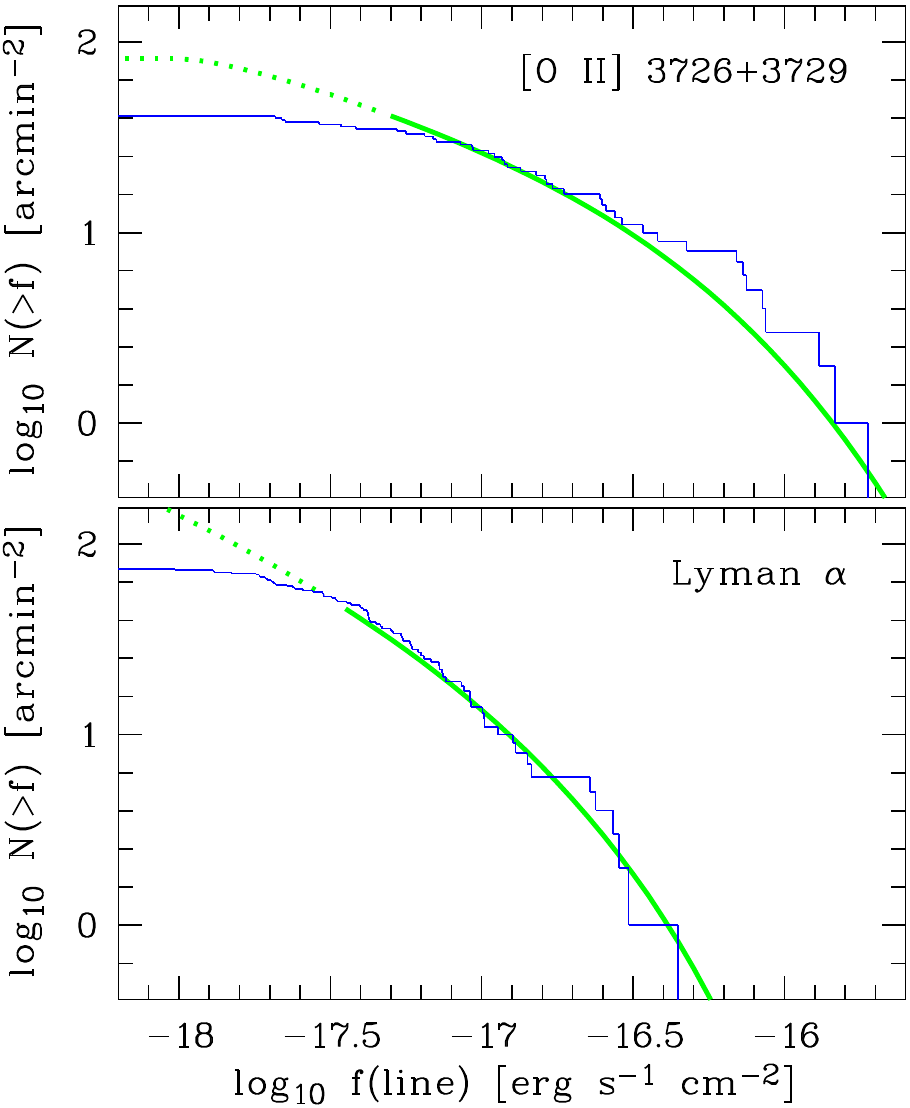} 
\caption{Cumulative number counts of emission line galaxies in the HDFS, as a function of line flux. Top panel: \oii\ emitters; bottom panel:\lya\ emitters. The green lines show the predictions for the relevant redshift ranges ($0.288 < z < 1.495$ and $2.95 < z < 6.65$, respectively), using a compilation of published luminosity functions as described in the text.} 
\label{fig:lw_nc}
\end{figure}

The resulting number counts of the two object classes, expressed as the cumulative number of objects per arcmin$^2$ brighter than a given flux, are depicted by the blue step functions in Fig.~\ref{fig:lw_nc}. Both panels show a relatively steep increase at bright fluxes which then flattens and eventually approaches a constant number density for line fluxes of $f\la 2\times 10^{-18}$~erg\,s$^{-1}$cm$^{-2}$, as no fainter objects are added to the sample. The detailed shapes of these curves depend on the emission line luminosity function and its evolution with redshift, but also on the selection functions which are different for the present samples of \lya\ and \oii\ emitters. A detailed discussion of this topic is beyond the scope of this paper. We note that at $5 \times 10^{-18}$~erg\,s$^{-1}$cm$^{-2}$, the numbers of detected \lya\ and\oii\ emitters are equal; at fainter flux levels, \lya\ emitters are more numerous than \oii\ emitters.

Contrasting the MUSE results for the HDFS to expectations from the
literature is not straightforward, as the relevant
surveys have very diverse redshift coverages. In the course of MUSE
science preparations, we combined several published results into 
analytic luminosity functions for both \lya\ and
\oii\ emitters, which we now compare with the actual
measurements. 
For the statistics of \oii\ emitters we used data by \citet{ly+2007}, \citet{takahashi+2007}, and \citet{Rauch+2008}, plus some guidance from the predictions for \ha\ emitters by \citet{geach+2010}. 
For the \lya\ emitters we adopted a luminosity function with non-evolving parameters following \citet{ouchi+2008}, with luminosity function parameters estimated by combining the \citet{ouchi+2008} results with those of \citet{gronwall+2007} and again \citet{Rauch+2008}. 
Note that these luminosity functions are just intended to be an overall summary of existing observations, and no attempt was made to account for possible tensions or inconsistencies between the different data sets.

Figure~\ref{fig:lw_nc} shows the predicted cumulative number counts based on these prescriptions as solid green curves (dotted where we consider these predictions as extrapolations). While the overall match appears highly satisfactory, cosmic variance is of course a strong effect in a field of this size, particularly at the bright end. We reiterate that any more detailed comparison with published luminosity functions would be premature. Nevertheless, it is reassuring to see that the present MUSE source catalogue for the HDFS gives number counts consistent with previous work.

\section{Spatially resolved kinematics}
\label{sect:resolved}

\begin{figure*} 
\resizebox{\hsize}{!}{\includegraphics{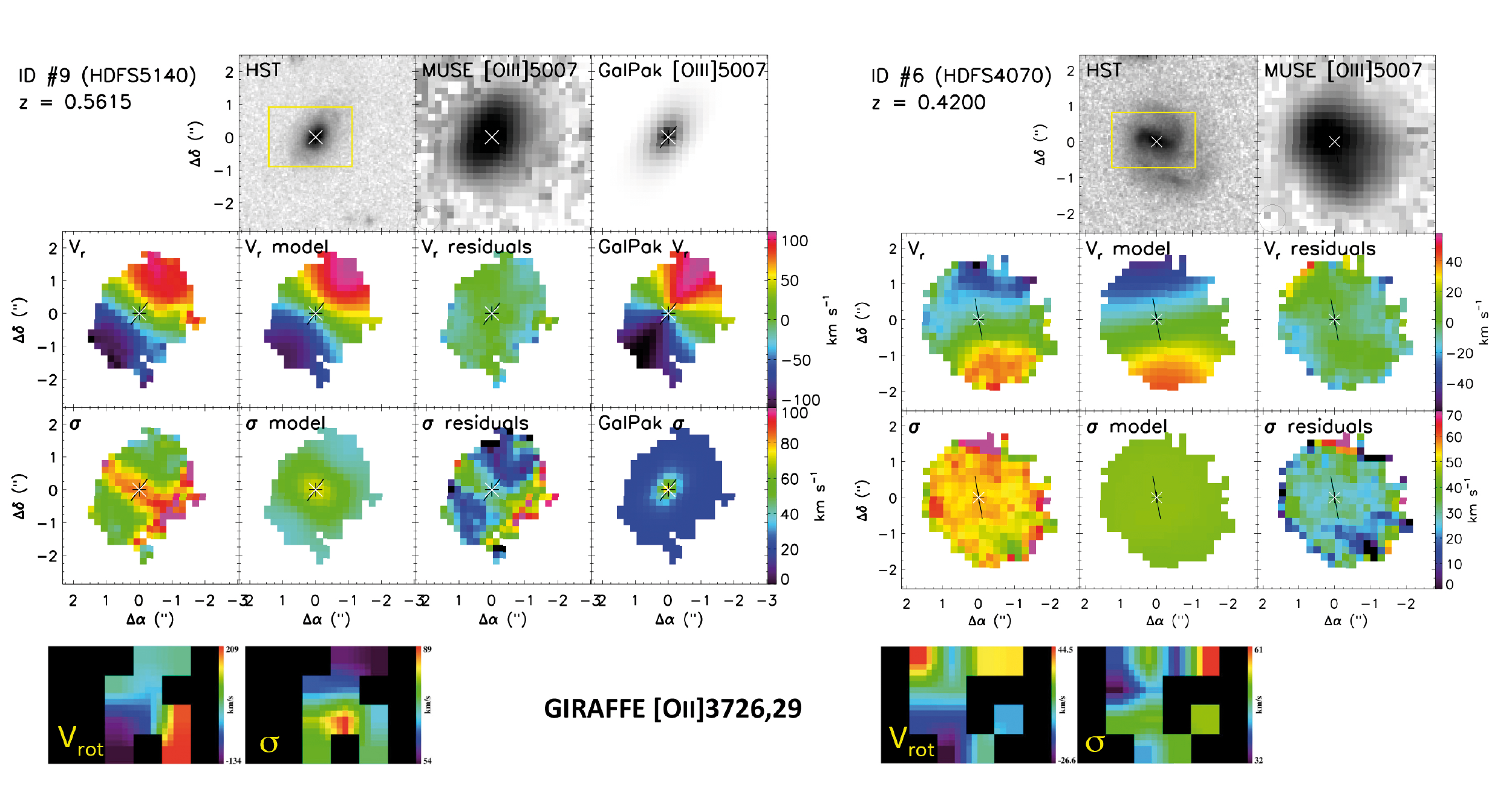}} 
\caption{Morpho-kinematics of HDFS5140 ({\it left}) and HDFS4070 ({\it right}). For each galaxy and from left to right. {\it Top}: HST/WFPC2 F814W image in log scale (the yellow rectangle shows the GIRAFFE FoV), the MUSE \oiiib\ flux map, and the deconvolved \oiiib\ flux map from GalPaK3D (for HDFS5140 only). {\it Middle}: MUSE observed velocity field from \hbeta\ and \oiii, velocity field of the 2D rotating disk model, residual velocity field, and deconvolved \oiiib\ velocity field from GalPaK3D (for HDFS5140 only). {\it Bottom}: MUSE observed velocity dispersion map from \hbeta\ and \oiii, velocity dispersion map deduced from the 2D velocity field model (beam-smearing effect and spectral PSF), deconvolved velocity dispersion map, and deconvolved \oiiib\ velocity dispersion from GalPaK3D (for HDFS5140 only). In each map, north is up and east is left. The center used for kinematical modeling is indicated as a white cross, the position angle is indicated by the black line which ends at the effective radius. For comparison, the published velocity field and dispersion map obtained with GIRAFFE for HDFS5140 \citep{Puech+06} and HDFS4070 \citep{Flores+06} are shown in the bottom row.}
\label{fig:kine}
\end{figure*}

To illustrate the power of MUSE for spatially-resolved studies of individual galaxies,
we derive the kinematics for  two galaxies with published spatially-resolved spectroscopy. These galaxies, namely \id{6} (hereafter HDFS4070) at $z=0.423$ and \id{9} (hereafter HDFS5140) at $z=0.5645$, were observed earlier with the GIRAFFE multi-IFU at the VLT as part of the IMAGES survey (\citealt{Flores+06}, \citealt{Puech+06}). These observations have a spectral resolution of $22-30$ \kms,  were taken with a seeing in the $0.35-0.8\arcsec$ range and used an integration time of 8 hours.These data have mainly been used to derive the ionized gas kinematics from the \oii\ emission lines, which was published in \citet{Flores+06} and \citet{Puech+06} for HDFS4070 and HDFS5140, respectively. As shown in the lower panels of Figure~\ref{fig:kine}, the velocity fields appear very perturbed. This led those authors to conclude that these galaxies show complex gas kinematics.

With the MUSE data in hand, we have the opportunity to revisit the ionized gas kinematics for these two galaxies and to compare the derived parameters with those obtained with GIRAFFE. To probe the gas kinematics, we make use of the three brightest emission lines available in the MUSE spectral range: the \oii\ doublet, \hbeta, and \oiii. 

The flux and velocity maps are presented in Fig.~\ref{fig:kine}. Rotating disk models, both in 2D (see \citealt{Epinat+12} for details) and with GalPaK3D (\citealt{Bouche+13}, 2014) are also shown. Compared to the GIRAFFE maps, the kinematics revealed by MUSE give a very different picture for these galaxies. 

The barred structure of HDFS4070 is clearly detected in the MUSE velocity field with its typical S-shape, whereas the velocity field of HDFS5140 is much more regular and typical of an
early-type spiral with a prominent bulge. The maximum rotation velocity derived for HDFS5140 is consistent between the 2D and 3D models ($\sim 140$ \kms), but is significantly lower than the value ($\sim 220$ \kms) obtained from the GIRAFFE data. 
In HDFS4070, the velocity dispersion is quite uniform over the disk whereas it is clearly peaked at the center of HDFS5140. Note, however, that there are some structures in the 
velocity dispersion residual map of this galaxy, with velocity dispersion of the order of $\sim 80-100$ \kms. Such a broad component, aligned along the south-west side of the minor axis of HDFS5140, could 
well be produced by superwind-driven shocks as revealed in a significant population of star-forming galaxies at $z\sim 2$ (see e.g. \citealt{Newman+12}). Note that outflows were already suspected in this galaxy from 
the GIRAFFE data, but not for the same reasons. \citet{Puech+06} argued that the velocity gradient of HDFS5140 is nearly perpendicular to its main optical axis, which is clearly not the case. 

The quality of the 2D MUSE maps uniquely enables reliable modeling  and interpretation of the internal physical properties of distant galaxies. A complete and extensive analysis of galaxy morpho-kinematics probed with MUSE in this HDFS field will be the scope of a follow-up paper (Contini, in prep).

\section{Summary and Conclusion}
The HDFS observations obtained during the last commissioning run of
MUSE demonstrate its capability to perform deep field
spectroscopy at a depth comparable to the HST deep photometry
imaging. 

\begin{figure} 
\resizebox{\hsize}{!}{\includegraphics{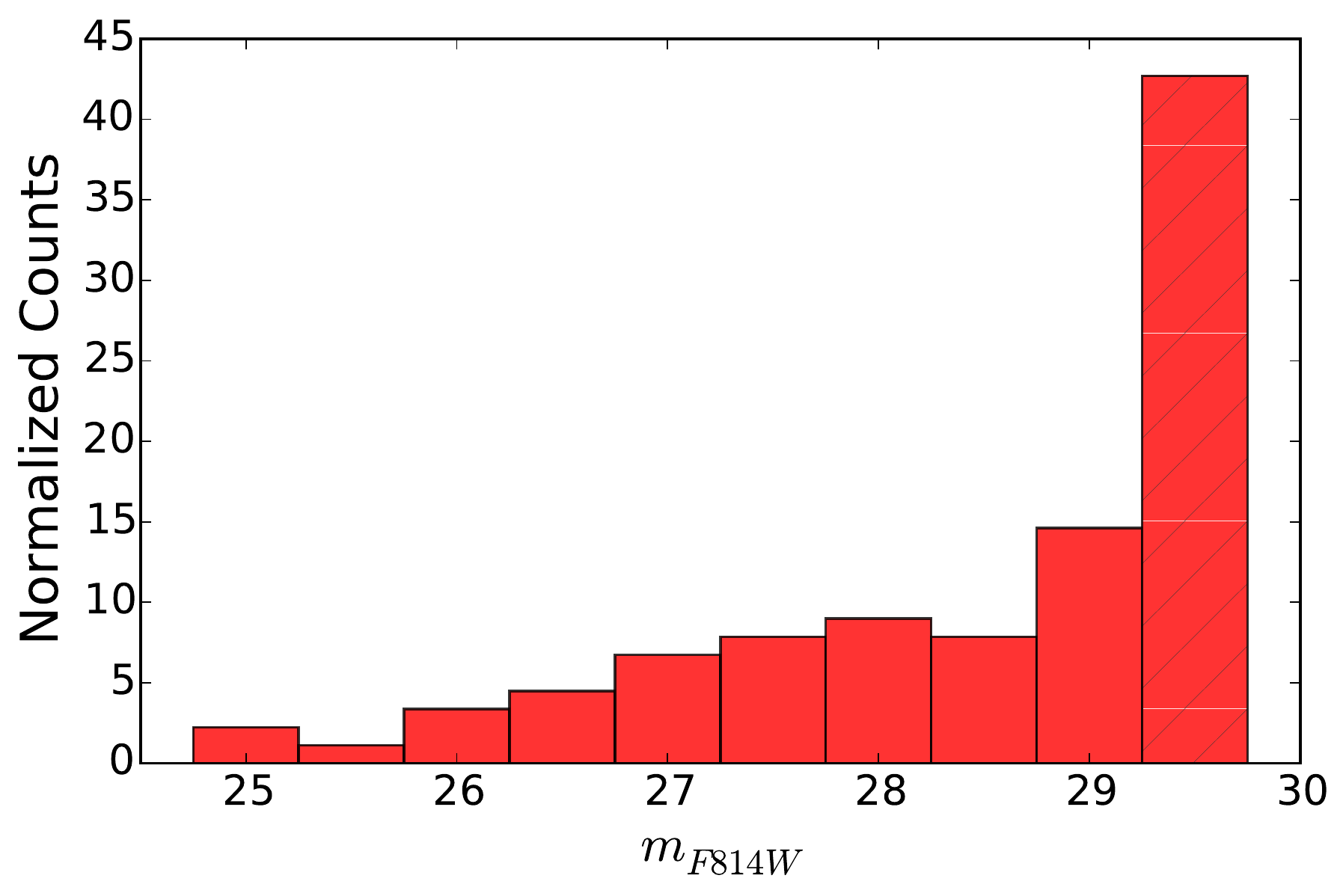}} 
\caption{The relative counts of MUSE HDFS 
  \lya\ emitters (in \%) versus UV continuum magnitude. 
  Note that the last bin of the histogram contains the 12 \lya\ emitters with magnitude $29.25 < $ \imag\ $< 29.75$ (taken from \citealt{Casertano+2000} catalog) and the 26 MUSE
  \lya\ emitters not detected by HST which have been arbitrarily assigned the
  average magnitude estimated from the stacked image \imag$=29.5$ (see Sect. \ref{sect:stackedlya})}
\label{fig:Lya-distri}
\end{figure}

In 27 hours, or the equivalent of 4 nights of observations with
overhead included, we have been able to get high quality spectra and
to measure precise redshifts for 189 sources (8 stars and 181 galaxies)
in the $1\, arcmin^2$ field of view of MUSE. This is to be compared
with the 18 spectroscopic redshifts that had been obtained before for (relatively bright) sources in the same area.  Among these 181 galaxies, we found 26 \lya\ emitters which were not even detected in the 
deep broad band WFPC2 images. The redshift distribution is different from the ones derived from deep multi-object spectroscopic surveys (e.g. \citealt{Lefevre+2014}) and extends to higher redshift. 
The galaxy redshifts revealed by MUSE extend to very low luminosities, as also
shown in Fig. \ref{fig:Lya-distri}.
Previous deep optical surveys like those of  \citet{Stark+2010} in the 
GOODS field did not push to such faint magnitudes.
\citeauthor{Stark+2010} obtained emission line redshifts to $AB=28.3$ and it can
be seen that MUSE can be used to derive  \lya\ fluxes and
redshifts to fainter limits. 
We achieved a completeness of 50\% of secure redshift identification
at \imag=26 and about 20\% at \imag=27-28. 

In the MUSE field of view, we have detected 17 groups with more than three members. The densest group
lies at $z=$1.284 and has nine members,
including 2 AGN and an interacting system showing a tidal tail. 
We also find three groups of \lya\ emitters at high redshifts ($z \sim $ 4.7,  4.9 and 5.7).

We have also investigated the capability of MUSE for spatially resolved spectroscopy of intermediate redshift galaxies. Thanks to its excellent spatial resolution, MUSE enables reliable modeling and interpretation of the internal physical properties of distant galaxies. Although the number of objects with relevant spatial information is limited to twenty in the MUSE HDFS field, this is still a large multiplex factor compared to single object pointing observations used e.g.\ for the MASSIV survey \citep{Contini+2012} with SINFONI at the VLT.

The MUSE observations of the HDFS are already the deepest spectroscopic survey ever performed in optical astronomy.  It is a unique data set and can be used for a large range
of follow-up studies. We release the reduced data world-wide, and make
the associated products public (see Appendix \ref{sect:public}).

The unique performance in ultra-deep spectroscopic fields achieved by MUSE is complementary to wide field surveys (e.g. \citealt{Lefevre+2014}) which can probe a much wider area on the sky but are limited to brighter objects and lower redshifts. With its capability to obtain high quality spectroscopic information of  the population of faint galaxies at high-redshift without having to perform any pre-selection, MUSE is highly complementary to the present (Hubble, ALMAs) and future (James Webb) ultra-deep surveys. As we have demonstrated using the HDFS observations, MUSE has in addition its own large potential for discoveries. MUSE is not limited to follow-up observations, but is also able to find large number of objects that are not (or barely) detected by broad band deep imaging.

These observations performed during the commissioning mark the end of the realization of the instrument and the start of the science exploitation. This is, however, only the first step. Further improvements of the data reduction process -- such as super flat-fields or optimal combination of exposures -- are still possible and will be implemented in the near future. We are also developing new methods for 3D source detections, optimized for the large size of the MUSE data cubes (\citealt{Paris+2013}, Herenz in prep., Meillier in prep, Cantalupo in prep).  With these new tools and an improved data reduction, we expect to  further extend the number of secure redshift identifications in the field. This will be the subject of a future public release for the HDFS field.

Another major milestone is expected in a few years, when the Adaptive Optics Facility will come in operations at UT4 \citep{Arsenault+2012}. The four laser guide stars, the deformable VLT secondary mirror and GALACSI, the module dedicated to MUSE \citep{Strobele+2012}, will improve the spatial resolution of MUSE up to 50\% without impacting its superb throughput and efficiency.

\begin{acknowledgements}
We thanks Laure Piqueras and Gerard Zins for their support with analysis and observation software. 
We also warmly thank Fernando Selman and all Paranal staff for their 
enthusiastic support of MUSE during the commissioning runs.
RB acknowledges support from the ERC advanced grant MUSICOS. 
JR, BC, VP acknowledges support from the ERC starting grant CALENDS.
TC, BG, AD acknowledges support from the ANR FOGHAR.
LM acknowledges support from the Lyon Institute of Origins under grant ANR-10-LABX-66.
SJL and KS acknowledge support from the Swiss National Science Foundation.

\end{acknowledgements}

\begin{appendix}
\section{Catalog}
\label{sect:catalog}

Table~\ref{sample-desc} lists the basic properties of the galaxies studied in the paper. All the sources from the \citet{Casertano+2000} catalogue that fall within the field of view of the final MUSE data cube are included as well as sources without an entry in the Casertano et al catalogue. The table is sorted by increasing apparent F814W magnitude from the catalogue with no particular ordering of the sources without a catalogue magnitude.

The first column gives a running number which is the one used for \id{XX} entries in the text. The subsequent two columns give the right ascenscion and declination from the MUSE observations, then follows the F814 SExtractor \texttt{BEST} magnitude from the Casertano et al catalogue and the $\mathrm{F606W}-\mathrm{F814W}$ colour from the \texttt{BEST} magnitudes. The redshift and its confidence follows thereafter, with the subsequent column indicating features identified in the spectrum. The final column gives the running number of the object in the Casertano et al catalogue. 

The second-to-last column, N$_\mathrm{exp}$, gives the median number of exposures going into the reduction of the region where the spectrum was extracted. Recall that the exposures are 30 minutes in duration, so a value of 40 corresponds to an exposure time of 20 hrs. Our redshift catalogue is naturally less complete where $\mathrm{N}_\mathrm{exp}<30$.

Table~\ref{sample-fluxes} gives the emission line fluxes measures off the 1D spectra produced by straight summation. In this table only the 181 sources with redshift $>0$ with redshift confidence $\ge 1$ are included. The procedure adopted for emission line flux measurements is given in \ref{sec:flux_measurements}.  Note that the fluxes have been corrected for Galactic reddening using the~\citet{SFD1998} dust maps using the reddening curve from ~\citet{ODonnell1994} for consistency with that work. 

We emphasise that there has been no attempt to correct these fluxes to true total fluxes. In particular these fluxes are generally lower than those used in section~\ref{subsec:lw_nc} which are aperture corrected. However the technique used in~\ref{subsec:lw_nc} is not suitable for all galaxies so we use the simpler approach here. Note also that the \lya\ fluxes have been measured by fitting a Gaussian to an asymmetric line so are likely to give sub-optimal flux measurements. Finally observe also that flux measurements are provided for all lines within the wavelength range of the spectrum, regardless of whether they were reliably detected or not.

Table~\ref{table:specz} gives the list of the redshift comparison between MUSE and published spectroscopic redshifts discussed in Section~\ref{subsec:zspec}.

\include{zcomp-table}

Note: catalogs will be made public once the paper is accepted for publication

\section{Public data release}
\label{sect:public}
In addition to the source catalog, we release the reduced data cube and associated files. We also deliver spectra and reconstructed images in the main emission lines for all the catalog sources. All this material is available at http://muse-vlt.eu/science.

Note: public data will be made public once the paper is accepted for publication
\end{appendix}

\bibliographystyle{aa}
\bibliography{hdfs-biblio.bib}

\end{document}

%% file: zcomp-table.tex
\onltab{
\begin{table*}
\label{table:specz}
\caption{Redshift Comparison}
\begin{tabular}{llcclllcl}
\hline\hline
\bf{id} & \bf{Object} & \bf{ra} & \bf{dec} & \bf{$m_{F814W}$} & \bf{$z_{MUSE}$} & \bf{$z_{lit}$}& \bf{Literature} & \bf{Comments} \\
2 & HDFS J223258.30-603351.7 & 22:32:58.30 & $-$60:33:51.66 & 21.3571 & 0.0 & 0.7063 & G06 & star \\
3 & HDFS J223253.74-603337.6 & 22:32:53.74 & $-$60:33:37.57 & 21.521 & 0.564497 & 0.5645 & S03 & --- \\
4 & HDFS J223252.14-603359.6 & 22:32:52.14 & $-$60:33:59.56 & 21.778 & 0.564616 & 0.5646 & S03 & --- \\
5 & HDFS J223252.24-603402.7 & 22:32:52.24 & $-$60:34:02.75 & 21.9726 & 0.580445 & 0.5804 & S03 & --- \\
6 & HDFS J223258.22-603331.6 & 22:32:58.21 & $-$60:33:31.64 & 21.9835 & 0.423033 & 0.4229 & S03 & Figure~\ref{fig:kine} \\
7 & HDFS J223259.43-603339.8 & 22:32:59.43 & $-$60:33:39.82 & 21.9948 & 0.464372 & 0.4644 & S03; G06 & --- \\
9 & HDFS J223256.08-603414.2 & 22:32:56.08 & $-$60:34:14.17 & 22.0849 & 0.56453 & 0.5645 & S03 & --- \\
10 & HDFS J223253.03-603328.5 & 22:32:53.03 & $-$60:33:28.53 & 22.5629 & 1.284517 & 1.27 & R05 & --- \\
13 & HDFS J223252.16-603323.9 & 22:32:52.16 & $-$60:33:23.92 & 22.8323 & 1.290184 & 1.293 & R05 & Figure~\ref{fig:spec_ID_13} \\
15 & HDFS J223252.88-603317.1 & 22:32:52.88 & $-$60:33:17.12 & 22.8565 & 1.284184 & 1.284 & W09 & --- \\
16 & HDFS J223255.24-603407.5 & 22:32:55.25 & $-$60:34:07.50 & 22.8682 & 0.465384 & 0.4656 & S03 & --- \\
17 & HDFS J223255.87-603317.8 & 22:32:55.87 & $-$60:33:17.75 & 22.926 & 0.581722 & 0.5817 & S03 & --- \\
20 & HDFS J223257.90-603349.1 & 22:32:57.90 & $-$60:33:49.12 & 23.0644 & 0.428094 & 0.428 & S03 & --- \\
23 & HDFS J223255.75-603333.8 & 22:32:55.74 & $-$60:33:33.75 & 23.4153 & 0.564872 & 0.5649 & S03; I05 & --- \\
41 & HDFS J223254.17-603409.1 & 22:32:54.18 & $-$60:34:08.94 & 24.571 & 2.4061 & 2.412 & I05; W09 & --- \\
43 & HDFS J223252.03-603342.6 & 22:32:52.03 & $-$60:33:42.59 & 24.6219 & 3.29254 & 3.295 & I05; W09 & Figure~\ref{fig:spec_ID_43} \\
55 & HDFS J223253.12-603320.3 & 22:32:53.11 & $-$60:33:20.25 & 24.9427 & 2.67 & 2.67 & I05; W09 & --- \\
87 & HDFS J223254.87-603342.2 & 22:32:54.86 & $-$60:33:42.12 & 25.7363 & 2.67 & 2.676 & W09 & --- \\
\hline
\end{tabular}
\tablebib{
 \citet[][S03]{Sawicki+2003}, \citet[][R05]{Rigopoulou+2005}, \citet[][I05]{Iwata+2005}, \citet[][G06]{Glazebrook+2006}, and \citet[][W09]{Wuyts+2009}.}
\end{table*}
}